%% file: scfe2020_v02.tex
\documentclass{svjour3}                     
\usepackage{graphicx,xcolor,natbib,hyperref}
\usepackage{latexsym,amssymb}
\usepackage{soul}

\usepackage[normalem]{ulem} 
\def\msun{\ifmmode M_\odot \else $M_\odot$ \fi}
\def\rsun{\ifmmode R_\odot \else R$_\odot$ \fi}
\def\lsun{\ifmmode L_\odot \else L$_\odot$ \fi}
\def\e{\ifmmode ^{-1} \else $^{-1}$ \fi}
\def\trh{\ifmmode \tau_{\rm rh} \else $\tau_{\rm rh}$ \fi}
\def\rh{\ifmmode r_{\rm h} \else $r_{\rm h}$\fi}
\def\vms{\ifmmode \langle v^2\rangle \else $\langle v^2\rangle$\fi}
\def\rhoh{\ifmmode \rho_{\rm h} \else $\rho_{\rm h}$\fi}
\def\Edotbin{\ifmmode \dot{E}_{\rm bin}\else $\dot{E}_{\rm bin}$\fi}
\def\Eext{\ifmmode E_{\rm ext}\else ${E}_{\rm ext}$\fi}
\def\Edotext{\ifmmode \dot{E}_{\rm ext}\else $\dot{E}_{\rm ext}$\fi}
\def\myr{\ifmmode {\rm Myr}\else Myr$\fi}
\def\mmax{\ifmmode m_{\rm max}\else $m_{\rm max}$\fi}

\def\CCquestion#1{{\color{magenta} \sf #1}}
\def\CC#1{{\color{blue} \sf #1}}

\newcommand{\HII}{H\,{\sc ii}}
\newcommand{\M}{{\cal M}}

\newcommand{\revI}[1]{{#1}}

\input{def}
\begin{document}

\title{The Physics of Star Cluster Formation and Evolution 
}


\author{Martin~G.~H.~Krause \and
        Stella~S.~R.~Offner \and 
        Corinne~Charbonnel \and
        Mark~Gieles \and
        Ralf~S.~Klessen \and
        Enrique~V\'azquez-Semadeni  \and 
        Javier Ballesteros-Paredes
        Philipp Girichidis \and
        J.~M.~Diederik Kruijssen \and 
        Jacob~L.~Ward \and
        Hans Zinnecker 
\institute{Martin G. H. Krause \at
              Centre for Astrophysics Research, 
              School of Physics, Astronomy and Mathematics, 
              University of Hertfordshire, College Lane,
              Hatfield, Hertfordshire AL10 9AB, UK \\
              \email{M.G.H.Krause@herts.ac.uk}           
              \and
            Stella S. R. Offner \at
              Department of Astronomy,
              The University of Texas,
              Austin TX, 78712, U.S.A.  
            \and
            Corinne Charbonnel \at
            Department of Astronomy, University of Geneva, Chemin de Pegase 51, 1290 Versoix, Switzerland; 
            IRAP, CNRS \& Univ. of Toulouse, 14, av.E.Belin, 31400 Toulouse, France
              \and
            Mark Gieles \at
             Institut de Ci\`{e}ncies del Cosmos (ICCUB-IEEC), Universitat de Barcelona, Mart\'{i} i Franqu\`{e}s 1, 08028 Barcelona, Spain;
             ICREA, Pg. Lluis Companys 23, 08010 Barcelona, Spain\\           
            \and
            Ralf S.\ Klessen \at
            Universit\"{a}t Heidelberg, Zentrum f\"{u}r Astronomie, Institut f\"{u}r Theoretische Astrophysik,  Albert-Ueberle-Str. 2, 69120 Heidelberg, Germany
                       \and
            Enrique V\'azquez-Semadeni \at
              Instituto de Radioastronom\'ia y Astrof\'isica,
              Universidad Nacional Aut\'onoma de M\'ex\'ico,
              Campus Morelia, Apdo. Postal 3-72, Morelia 58089, M\'exico
            \and
           Javier Ballesteros-Paredes \at
              Instituto de Radioastronom\'ia y Astrof\'isica,
              Universidad Nacional Aut\'onoma de M\'ex\'ico,
              Campus Morelia, Apdo. Postal 3-72, Morelia 58089, M\'exico
            \and
            Philipp Girichidis \at
            Leibniz-Institut f\"{u}r Astrophysik (AIP), An der Sternwarte 16, 14482 Potsdam, Germany
             \and
           J.~M.~Diederik~Kruijssen \at
              Astronomisches Rechen-Institut,
              Zentrum f\"ur Astronomie der Universit\"at Heidelberg,
              M\"onchhofstra\ss{}e 12-14, 69120 Heidelberg, Germany
             \and
           Jacob L.~Ward \at
              Astronomisches Rechen-Institut,
              Zentrum f\"ur Astronomie der Universit\"at Heidelberg,
              M\"onchhofstra\ss{}e 12-14, 69120 Heidelberg, Germany
            \and
            Hans Zinnecker \at 
            Nucleo de Astroquimica y Astrofisica, Universidad Autonoma de Chile, Avda Pedro de Valdivia 425, Providencia, Santiago de Chile, Chile}
}
\date{Received: 31 Jan 2020 / Accepted: date}

\maketitle
\begin{abstract}
Star clusters form in dense, hierarchically collapsing gas clouds. Bulk kinetic energy is transformed to turbulence with stars forming from cores fed by filaments. In the most compact regions, stellar feedback is least effective in removing the gas and stars may form very efficiently. These are also the regions where, in high-mass clusters, ejecta from some kind of high-mass stars are effectively captured during the formation phase of some of the low mass stars and effectively channeled into the latter to form multiple populations. Star formation epochs in star clusters are generally set by gas flows that determine the abundance of gas in the cluster. We argue that there is likely only one star formation epoch after which clusters remain essentially clear of gas by cluster winds. Collisional dynamics is important in this phase leading to core collapse, expansion and eventual dispersion of every cluster. We review recent developments in the field with a focus on theoretical work. 
\end{abstract}

\section{Star clusters: more than a collection of stars}\label{sec:intro}
%
Star clusters have caught human attention since ancient times,
as evidenced for example by depictions of the Pleiades on cave walls
and the Nebra Disk \citep{Rap01,Mozel03}. They continue to be
a fascinating topic today, thanks to new and puzzling observations
challenging the theoretical models.

\begin{figure}\centering
  \includegraphics[width=1.0\textwidth]{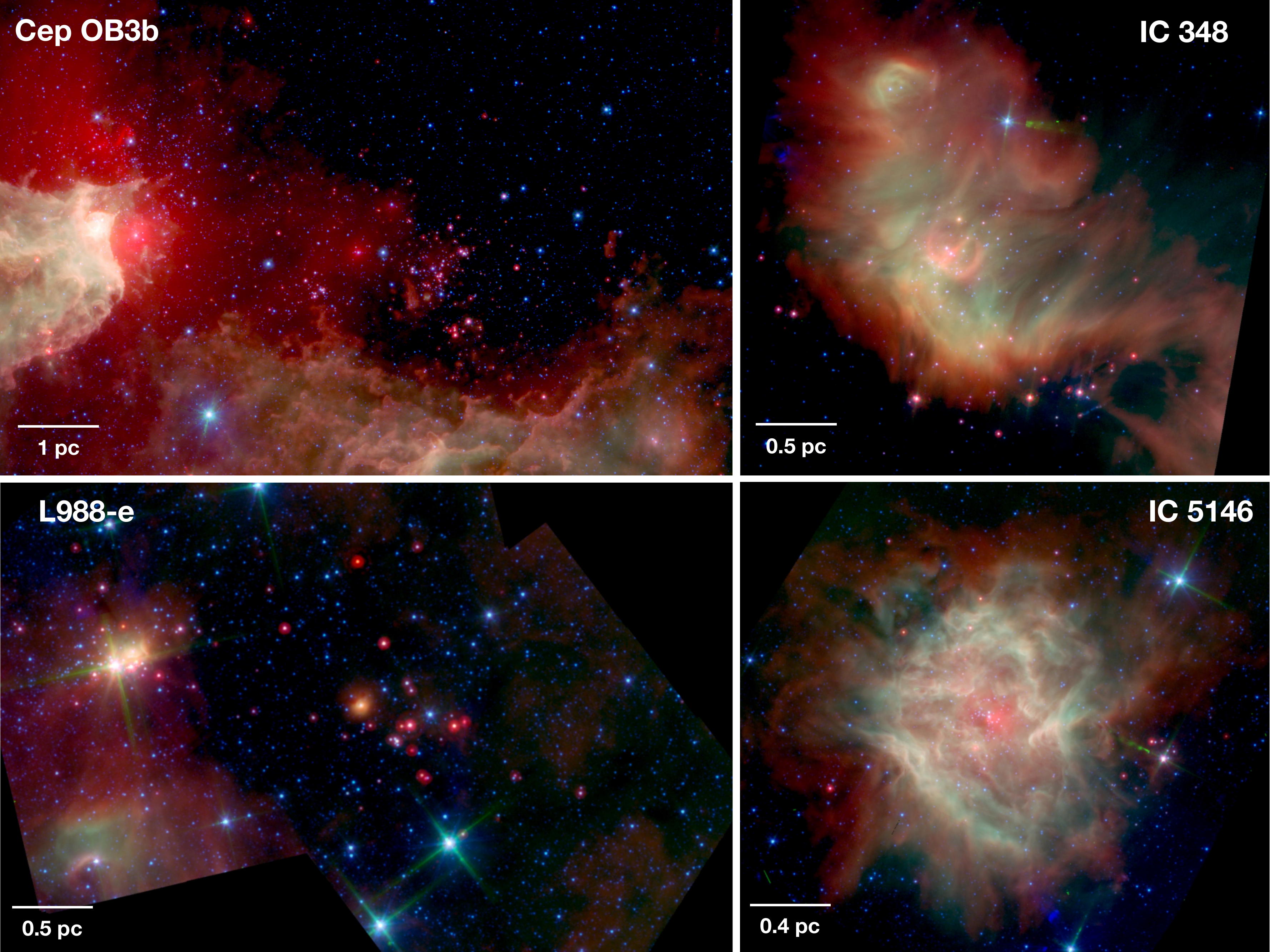}
\caption{Three-color Spitzer images (3.6 (blue), 5.8 (green), and 24 $\mu$m (red)) of young, nearby ($d<1000$\,pc) clusters. Cluster source catalogs are contained in the Spitzer Extended Solar Neighbourhood Archive (SESNA, Gutermuth et al.~in prep). Top left: Young (4-5 Myr), massive cluster Cep OB3b, which is part of the Cep OB3 molecular cloud complex \citep{Gutermuth+2011,Allen+2012}. Top right: IC 348 cluster (2-3 Myr), which is forming in a sub-region of the Perseus molecular cloud \citep{2009ApJS..184...18G}. Bottom left: Extended field containing exposed cluster L988-e (courtesy of R. Gutermuth). Bottom right: Small, dense cluster IC 5146, where protostars are forming around a bright PAH emission bubble \citep{2009ApJS..184...18G}.}
\label{fig:obs_clusters}       
\end{figure}

Spitzer has traced the dense gas in a number of nearby young clusters
(Fig~\ref{fig:obs_clusters}) and shown its connection to young stellar objects
\citep[e.g.,][]{Gutermuth+2011}. Thanks to GAIA \citep{GAIA_DR2}, we now know the kinematics of many clusters 
on a star-by-star basis \citep[e.g.,][]{WardKruijssen2018,Karnea19,Kuhnea19}. 
Chemistry is traced by spectroscopic and photometric surveys \citep{BL18,Grattea19};
cluster winds have been detected in spaceborne X-ray observations \citep{KNM11} and young
super star clusters show evidence of MASER emission \citep{Gorskea19}.

These observations place strong constraints on theoretical modelling.
The latter has been typically attempted from different angles with a 
view on explaining a particular subset of observations. A simulation
that includes gas dynamics, stellar dynamics and chemistry to sufficient
accuracy and from cloud collapse to cluster dispersal remains beyond reach
for the foreseeable future. Approaches that focus on each aspect separately, 
or combine some aspects making some approximations therefore have to form
the basis of our understanding of stellar clusters.

This review aims to provide an overview of the different theoretical 
approaches, puts them in context with each other, and aims to paint a 
comprehensive and coherent picture of the physics of star cluster
formation and evolution.
We are not aware of past projects with such an ambition, but 
previous reviews that have significant overlap with the present one 
include \citet{MLK04,ZY07,PZMcMG10,GCB12,Renzini13,Kruij14,Krumholz14,Longmea14,Charb16,Klessen2016,BL18,Grattea19} and \citet{Krumhea19}.
%
After defining star clusters in \S\ref{sec:def},
we first review the onset of star formation in 
molecular clouds (\S\ref{sec:frag}) and the 
formation of stars in clusters (\S\ref{sec:sf}).
In this formation period, physical processes and multiple scales are coupled. From the end of the star formation epoch, gas dynamics (\S\ref{sec:gd}), stellar dynamics (\S\ref{sec:nbody}) and nucleosynthesis (\S\ref{sec:nucsy}) evolve independently. Each section includes, however, links to the other fields.
In particular, the chemistry in the predominantly old, multiple population clusters, discussed in \S\ref{sec:nucsy}, refers back to the formation epoch, where all the different processes are coupled.
We conclude with a summary
and outlook in \S\ref{sec:syn}.

\section{What is a star cluster?}\label{sec:def}
We 
adopt the ontological definition that a star cluster is 
a gravitationally bound group of stars inside a closed tidal surface if this volume is
\begin{enumerate}
\item not dark matter-dominated and
\item contains at least 12 stars. 
\end{enumerate}
The first condition distinguishes star clusters from galaxies.
The second one from multiple star systems. This definition
essentially follows \citeauthor{Krumhea19} (\citeyear{Krumhea19}, though we do not distinguish here between
different overdensities required in different environments). 
Groups of stars that are not gravitationally
bound are called associations \citep[ and Adamo et al. 2020, in prep.]{1964ARA&A...2..213B,GiPo11,Gouliermis2018}. For the Milky Way,
bound star clusters have been subdivided into open clusters
in the disc and globular clusters associated with the bulge and halo. 
Open clusters are generally young ($\lesssim 1$~Gyr) and have 
low mass ($\lesssim 10^5 M_\odot$) while globular clusters are generally old 
($> 1$~Gyr) \revI{and} massive ($\gtrsim 10^4 M_\odot$)\revI{, quite typically} survivors from 
the early Universe, representing the relics of star formation at high redshift.
\revI{In fact, the oldest globular clusters in the Milky Way have a likely age 
$>13$~Gyr and provide an important constraint for the age of the Universe 
\citep{2003Sci...299...65K,2017ApJ...838..162O}.}
The distinction \revI{between open and globular clusters}
happens to correspond closely to a fundamental distinction in 
photometric properties and chemical abundance patterns: 
Open clusters
are mostly single population clusters with a single main sequence
in the colour-magnitude diagram, while almost all globular clusters have 
multiple main sequences and strong star to star variations 
in light-element abundances, i.e., multiple stellar populations.
A more useful classification of star clusters is therefore 
between single and multiple population clusters \citep{Carea10a,BL18}.

\section{The onset of star formation in molecular clouds} \label{sec:frag}
    
Star clusters form from molecular clouds, which are the densest regions in the interstellar medium, and consist mostly of molecular hydrogen and several other molecules, which are used as tracers for observing these regions and their substructure. Molecular clouds range in mass from $\sim 10^3$ to $\sim 10^7 \msun$, and have extremely complex {\it hierarchical} (or fractal) morphologies \citep{1996ApJ...471..816E}, with the densest regions embedded in larger, lower-density ones, and so on \citep[e.g.,] [] {BW99}. It has been suggested that the internal structure and dynamics of molecular clouds is instrumental in determining the early structure and kinematics of star clusters \citep[e.g.,] [hereafter VS17] {Klessen+00,Klessen01a,Klessen01b,OffnerEtAl2009,Kruijea12a,GirichidisEtAl2012b, VS+17}.


\subsection{The Gravoturbulent (GT) scenario} \label{sec:GT}

Molecular clouds are known to have internal supersonic non-thermal motions \citep{Wilson+70}, which follow a relation between the observed linewidth and the spatial scale \citep[Larson's relation][]{Larson81, Hennebelle2012}, although with substantial scatter \citep{BallesterosParedes+11a,Miville+2017}. 
These supersonic motions were originally interpreted as large-scale radial motions, likely to originate from global collapse \citep{Liszt+74, GK74}. However, this interpretation was soon rejected because, as it was argued, it would lead to excessively large star formation rates (SFRs) and should produce systematic velocity differences (i.e., red or blue line shifts) between emission lines produced by HII regions located at the centers of the clouds and absorption lines produced at the outskirts of the clouds. Since such shifts were not observed, the supersonic motions were then interpreted as {\it small-scale} supersonic turbulence that produces a turbulent pressure capable of supporting the clouds against their self-gravity \citep{ZP74, ZE74}. The requirement for the motions to be confined to small scales was necessary in order to avoid the generation of the unobserved line shifts and to produce an isotropic pressure that could support the clouds.

 Since then, the prevailing paradigm for molecular clouds is that they are supported by some agent against their self-gravity, typically turbulence and/or magnetic fields. These are invoked in part to explain the observed star-formation efficiencies per free-fall time, $\epsff$ -- the fraction of gas mass converted to stars over a free fall time -- of $\sim 1\%$  for most giant molecular clouds (GMCs) \citep{ZE74,Krumhea19}. 
Since turbulence is known to dissipate rapidly, typically in a crossing time, it was first proposed that the motions consisted of Alfv\'enic turbulence, because Alfv\'en waves were thought to be less dissipative than shocks \citep[e.g., ] [] {Shu+87}. However, subsequent numerical simulations of MHD turbulence showed that it dissipates as rapidly as hydrodynamic turbulence \citep{ML+98, Stone+98, PN99}, implying that constant driving of the turbulence must be present to maintain it. In this {\it gravoturbulent} (GT) scenario \citep[e.g., ] [] {Klessen+00, VS+03, MLK04}, the clouds are supported globally by the pressure of the (continuously driven) supersonic, small-scale, isotropic turbulence, while locally, shocks are produced that in turn generate density fluctuations (sheets, filaments, and clumps), which may locally become Jeans unstable and collapse. Moreover, if magnetised turbulence provides support such that clouds are neither dispersing nor globally collapsing, it is assumed that the clouds are in approximate virial equilibrium between turbulence and self-gravity \citep{KMM06,Goldbaum+2011}. This assumption is consistent with observations \citep[e.g., ] [] {Larson81,Heyer+09}.
In particular, the \citet{Larson81} linewidth-size relation observed in molecular clouds is interpreted as the manifestation of the energy spectrum, $E(k) \propto k^{-2}$, corresponding to strongly compressible, highly supersonic turbulence.

\subsection{The Global Hierarchical Collapse (GHC) scenario} \label{sec:GHC}

On the other hand, there is evidence that the process of formation of the molecular clouds is important for their subsequent dynamical evolution. The clouds seem to form by accreting tenuous ($n \sim 10\, \pcc$) atomic gas, which often appears gravitationally bound to the molecular gas it surrounds \citep{Fukui2009}. Moreover, molecular clouds exhibit a hierarchical structure, so that their internal dynamics are governed by very similar processes. On smaller scales,
star-forming cores accrete material from the scale of their parent clumps \citep[i.e., the cores are said to be {\it clump-fed};] [] {Liu+15, Yuan+18}, and longitudinal, multi-parsec scale flows are routinely observed along filamentary clouds, which feed the main cores (or {\it hubs}) within the filaments \citep[e.g.,] [] {Myers2009, Schneider+10, Kirk13, Peretto+2014, 2016A&A...585A.149W,Hacar+17, Chen+2019}. Additionally, numerical simulations of the formation and evolution of cold, dense atomic clouds from large-scale compressions in the warm, diffuse gas also suggest that the clouds engage into global, hierarchical collapse \citep[GHC;] [] {VS+19} soon after they reach their thermal Jeans mass \citep{VS+07, VS+09, Heitschea08}.  In what follows, we focus on this scenario, as it provides a direct link between the processes occurring in the gas during the collapse and the structural properties of the resulting stellar cluster(s).

\subsubsection{Onset of large-scale gravitational contraction and turbulence generation}

The clouds are expected to rapidly reach and exceed their thermal Jeans mass because the Jeans mass in the dense, cold gas is $\sim 10^4$ times smaller than in the diffuse, warm gas \citep{GomVaz14} and simulations indicate that the clouds actively accrete from their diffuse environment \citep{BP+99, Hartmann+01, VS+06, HeiHart08, Banerjee+09, Heiner+15, Wareing+19}. This accretion implies that the clouds generally grow in mass, allowing them to become magnetically supercritical (i.e., unsupported by the magnetic field), gravitationally unstable, and molecular at roughly the same column density ($\sim 10^{21}$ \psc) for solar-neighbourhood pressures and metallicities \citep{Hartmann+01, Heitsch+09, VS+11, Heiner+15}.

Simulations of the self-consistent formation and evolution of clouds by converging streams of diffuse gas \citep[e.g.,] [] {Heitsch+05, Heitsch+06, AudHen05, AudHen10, VS+06, VS+07, Hennea08, Banerjee+09}
show that the very formation process of the cloud causes the generation of moderately supersonic (with respect to the sound speed in the cold, dense gas) turbulence by the combined action of various instabilities, such as the nonlinear thin-shell instability \citep{Vish94}, thermal instability \citep{Field65} and Kelvin-Helmholtz instability (see \citeauthor{Heitsch+06}, \citeyear{Heitsch+06}, and \citeauthor{KH2010}, \citeyear{KH2010}, for  further discussions). 
Similar effects have also been shown for shells
of expanding bubbles \citep{Krausea13a}.
The turbulence generates nonlinear density fluctuations in which the free-fall time $\tau_{\rm ff} = \sqrt{3 \pi / (32 G \rho)}$ is significantly shorter than the average in the cloud. 

The energy in the turbulent motions generated by the instabilities, which \revI{are} only moderately supersonic with respect to the cold gas, and subsonic with respect to the warm gas, quickly becomes overwhelmed by the gravitational energy of the whole cloud (actually, a cloud complex), which then begins to undergo global gravitational contraction. In the GHC scenario, thus, the apparent near-virial state of molecular clouds and their substructures is not due to turbulent support, but rather to the infall motions driven by the self-gravity \citep{BallesterosParedes+11a}. It should be noted, however, that the infall is highly chaotic and so a truly random (turbulent) component is in fact maintained by the collapse \citep{KH10, RG12, MC15, LiGX18}, although it is apparently not able to significantly delay the collapse, possibly due to the rapid dissipation. More experiments are needed to clarify exactly how much turbulence is generated by the collapse, especially in the context of the formation of the first stars, where energy loss via radiative cooling is suppressed due to the low metallicity of the gas \citep {Sur2012,Latif2013,Schoberb,Bovino2013,Federrath2014,Schober2015, Klessen2018}.



The presence of turbulent density fluctuations with nonlinear amplitudes, together with the generally amorphous and flattened or filamentary shape of the clouds, has the important implication that realistic collapse is far from homologous (uniform spherical configurations, all material in the sphere reaching the center at the same time). It is well known that already in non-uniform spherical configurations (``cores") with centrally-peaked radial density profiles, the central, densest parts terminate their collapse (i.e., reach protostellar densities) earlier than the outer parts, and then the rest of the material, which was initially at lower densities, continues to accrete onto the previously collapsed material \citep[e.g.,] [] {Larson1969, Penston1969, Shu1977, Hunter77, WS85, FC93, MS13, Keto+15, Naranjo+15}. 
In a turbulent system, the nonlinear density fluctuations have free-fall times significantly shorter than that of the whole cloud, and so they can collapse faster, as soon as they become locally gravitationally unstable \citep[compare][]{VS+19}. 

Under this regime, the cloud evolves towards containing a large number of thermal Jeans masses, in agreement with the observation that molecular clouds typically have masses $\Mc$ upwards of $10^3 \msun$ \citep[e.g., ] [and references therein] {MLK04}. 
Thus, the cloud becomes a system of {\it collapses within collapses}, with an ever-larger hierarchy of collapsing scales, each one accreting from the next larger scale \citep{VS+19}. This is a mass cascade, in some senses similar to the turbulent energy cascade \citep{Field+08}. This is also essentially Hoyle's fragmentation \citep{Hoyle53}, except with nonlinear density fluctuations and non-spherical geometry of the clumps \citep{VS+19}. Also, it can be considered as an extension of the {\it competitive accretion} scenario \citep{Bonnell01, BonnellBate2006}, with the accretion extending to  cloud scales ($\sim 10$ parsecs or more), and with the added ingredient that a whole hierarchy of chaotic, gravitational contraction flows is present.

\subsubsection{Filament formation and filamentary accretion}

At sufficiently advanced stages of a cloud's evolution, when its mass $\Mc$ is much larger than the Jeans mass, it must behave essentially as a pressureless collapse, because precisely the meaning of $\Mc \gg \MJ$ is that the gravitational energy overwhelms the internal energy of the cloud. But it is known that pressureless collapse amplifies anisotropies, so that a triaxial ellipsoid contracts first along its shortest dimension to form a sheet, and then an elliptical sheet contracts again along its shortest dimension to form a filament \citep{Lin+65}. Therefore, it is expected that multi-Jeans mass molecular clouds should evolve to develop filaments, which are actually akin to ``rivers" funnelling the mass from large to small scales \citep{GomVaz14}. This is consistent with the observation that dense molecular cloud cores appear as ``hubs" at the intersection of filaments \citep[e.g., ] [] {Myers09}, with the filaments feeding material to the hubs \citep[e.g., ] [] {Schneider+10, Sugitani+11, Kirk13, Peretto+2014, Chen+2019}. 

Since the majority ($\sim 90\%$) of pre- and protostellar cores in molecular clouds are located either in filaments or in hubs \citep{Konyves+15}, it follows that star formation is initiated already in the flows feeding the hubs. This mechanism was referred to as ``conveyor belt cluster formation" by  \citet{Longmea14} and in \citet{Krumhea19}, in opposition to ``monolithic cluster formation", in which the gas first collapses and subsequently forms stars in a centrally-concentrated cluster. The conveyor-belt mechanism is also observed in simulations of self-consistent cloud formation and evolution, in which the filaments form spontaneously by anisotropic gravitational contraction \citep{GomVaz14, VS+19}.

\subsubsection{Acceleration of star formation and delayed formation of massive stars}

Another expected consequence of the global collapse and continued accretion onto the star-forming hubs observed in the simulations is an acceleration of the star formation before massive stars form. This increase in the star formation rate (SFR) is routinely observed in simulations of cloud evolution \citep[e.g., ] [] {VS+10, VS+17, Hartmann+12, Colin+13, LeeE+15, LiPS+18}, and predicted by models of clouds dominated by gravity \citep[e.g., ] [] {ZA+12, ZV14, CC18}. Observational evidence of the acceleration is provided by, for example, a) the age histograms of young embedded clusters, which systematically show a maximum at either the smallest ages, or at a certain, relatively recent age, together with a tail of older stars, of ages up to several Myr \citep[e.g., ] [] {BP+99, PS00, PS02, HS06, DaRio+10}; b) a superlinear ($\sim t^2$) temporal dependence of the total number of stars formed at time $t$ in several young clusters \citep{CC18}.

\begin{figure}
    \centering
    \includegraphics[width = 0.49\textwidth] {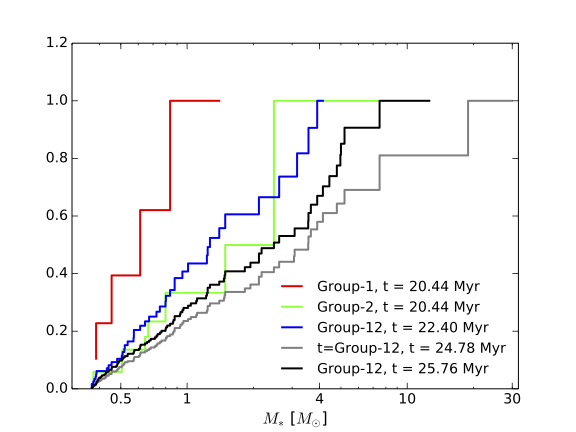}
    \includegraphics[width = 0.49\textwidth] {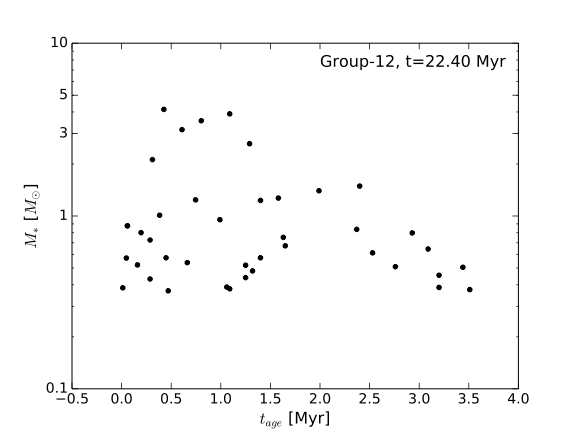}
    \includegraphics[width = 0.49\textwidth] {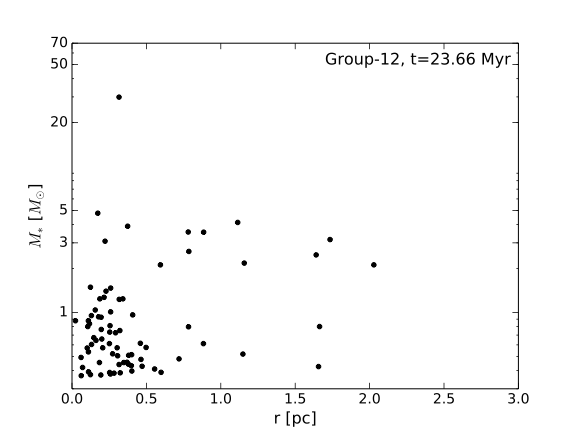}
    \includegraphics[width = 0.45\textwidth] {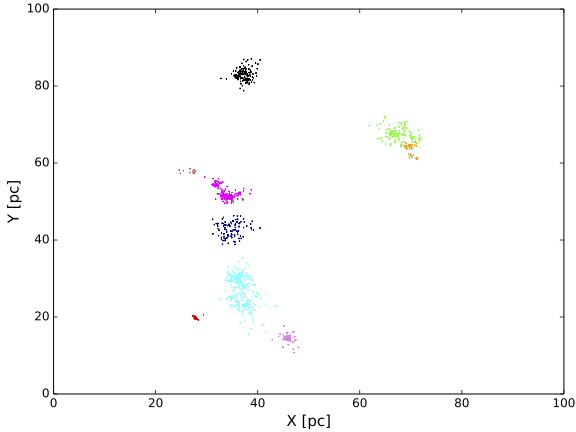}
    \caption{{\it Top left:} Normalised cumulative stellar mass histograms of star forming regions in a numerical simulation of GHC \citep{VS+17} at various times. As time proceeds, a larger fraction of the stars are seen to be massive, until feedback begins to disrupt the cloud. {\it Top right:} Mass {\it versus} age of the members of a stellar group in the same simulation, at time $t = 22.4$ Myr, corresponding to roughly 3.5 Myr after the onset of star formation in the region. {\it Bottom left:} Mass {\it versus} distance from center of mass of the group members at $t = 23.7$ Myr in the simulation ($\approx 4.7$ Myr after the onset of star formation. The more massive stars are seen to be located near the center of mass. {\it Bottom right:} Groups constituting the cluster in the simulation at time $t = 30.0$ Myr (11 Myr after the onset of SF), as identified by a friends-of-friends algorithm. \revI{Stars belonging to the same group have the same colours.} Figures from \citet{VS+17}.}
    \label{fig:age-mass_distribution}
\end{figure}

In the GHC scenario, the increase of the SFR during the early stages of star-forming regions is due to the growth in mass, density, and size of the regions due to accretion from their parent structures. The increase in density implies an increase of the SFR because a larger fraction of the mass is at densities high enough that their free-fall time is much shorter than that of the mean density of the parent structure \citep{ZV14}. But, additionally, the larger mass of the more evolved regions provides a larger mass reservoir, allowing for the formation of more massive stars. So, the star-forming regions evolve towards forming more massive stars, meaning that the formation of massive stars is delayed with respect to that of the first low mass stars, by several Myr in moderate-mass regions, according to the simulations \citep[] [top left panel of Fig.\ \ref{fig:age-mass_distribution}] {VS+09, VS+17}. Note that low-mass stars always form, but the maximum mass of the stars that can form is capped by the instantaneous mass of the hub where they form, and increases with time as long as the hub's mass increases. Eventually, however, the stellar feedback begins to erode the hub in the simulations, decreasing its density and mass, and also eroding the filamentary accretion flow, decreasing the maximum stellar mass that can form. A model for the development of the high-mass slope of the IMF based on the same principle, of the mass of the most massive star being bounded by the mass of the hub in which it forms, has been developed by \citet{Oey11}. This delayed formation of massive stars also implies that the age range of the massive stars is smaller (and they are younger) than that of the low-mass stars, which begin to form since the onset of the star formation activity in the region. Equivalently, the mass range of the younger stars is larger and extends to higher masses than that of the older stars (top right panel of Fig.\ \ref{fig:age-mass_distribution}).

\subsubsection{Mass and age radial gradients}

Star formation occurring in the filaments generally involves lower-mass cores, because they are themselves part of the flow falling onto the main hubs, which are the main accreting centers. That is, the stars formed in the filaments do so in secondary gravitational potential wells, while the hubs are the primary wells. Therefore, the more extended secondary star formation in the filaments generally produces lower-mass stars, and thus tends to produce a primordial mass segregation in the cluster (bottom left panel of Fig.\ \ref{fig:age-mass_distribution}), independent of any $N$-body processes that may occur afterwards
(\S\ref{sec:mseg}). 

Because the secondary star formation in the filaments, involves lower-mass stars that have been forming for a longer time, the median age of the more distant stars tends to be larger than that of those nearer the hub. In the simulation from Fig.~\ref{fig:age-mass_distribution} a median-age gradient of $\sim 1$ Myr pc$^{-1}$ is found \citep{Getman+18}, consistent with the gradient observed in the clusters in the MYStIX \citep{Feigelson+13} and SFiNCs \citep{Getman+17} star-forming region catalogs.

The hierarchical and filamentary structure of the collapse flow is imprinted on the structure of the cluster itself, which therefore adopts a self-similar, fractal-like spatial distribution, and retains traces of a filamentary morphology, as seen in the bottom right panel of Fig.\ \ref{fig:age-mass_distribution}, which shows the groups constituting the cluster in the simulation at time $t = 30.0$ Myr, as identified by a friends-of-friends algorithm. The groups are seen to be strung along a long filamentary structure, and also, when the linking parameter of the algorithm is varied, the number of identified groups varies, indicating a hierarchical structure \citep[see] [for further details] {VS+17}.

The above structural properties of the nascent cluster are blurred to some degree when the the massive stars lose their mass, the gas is cleared by the stellar feedback and the stars have time for $n$-body interactions (see below).  Therefore, these ``primordial" structural features are expected to be more prominent in younger clusters.

\section{The formation of stars in clusters}
\label{sec:sf}

 The formation and evolution of star clusters is a multi-scale process that depends in detail on the formation of the constituent individual stars.  In turn, forming stars influence the accretion and dynamics of their neighbours through stellar feedback, including winds, radiation and supernovae.

Stars accreting from a common gas clump or protostellar core may compete with one another for fuel, while accretion disk properties depend on the local ionising flux, which is set by the distribution of nearby massive stars. Ionising radiation, winds and jets from protostars interact with their own accretion streams, as well as those of other stars. Finally, there is the puzzling observation of multiple populations in clusters (\S\ref{sec:nucsy}), where massive star ejecta may affect the accretion flows of lower mass stars. Understanding how an individual star in a cluster grows by accretion of gas and why it reaches a particular mass is therefore inseparable from the larger cluster context.

During the earliest stages of accretion, stars are hidden from view, and direct probes of accretion, like stellar spectral lines, which are exploited to study accretion in T-Tauri stars, are unavailable. Instead, indirect evidence of accretion, such as protostellar outflows and luminosities must be used to reconstruct the magnitude and history of accretion. In this section, we first discuss the observational signatures of accretion and their implications. We then summarise a variety of theoretical models for protostellar accretion. We sub-divide these into three categories: models based on the properties of cores and/or filaments that host the protostar, models that focus on the protostar-disk relationship and models that depend on feedback and the larger protostellar environment. This division is mainly for convenience, since in practice, accretion is determined by a variety of nonlinear processes that span a broad range of times and physical scales. Finally, we discuss models for how and why accretion ultimately ceases, which is critical for understanding the accretion histories of individual stars as well as global properties such as the star formation efficiency, star formation rate and lifetime of molecular clouds.

\subsection{Observational Signatures of Accretion: Protostellar Luminosities, Outflows and Spectral Lines} 

Protostellar outflows are a direct byproduct of the accretion process (see Chapter Processes for more details). If a fixed fraction of accreting material is flung outwards in an outflow, then in principle by measuring the outflow mass flux it is possible to reverse engineer the accretion rate and history. Observations of protostellar outflows suggest protostellar accretion rates of  $10^{-4}-10^{-9}\,\msun$\,yr\e, where younger or more massive protostars have higher inferred accretion rates \citep{Bally2016}. 

Outflow morphology also gives important insights in the accretion process. Outflows and jets (highly-collimated flows, usually observed in optical emission), frequently exhibit regularly spaced clumps along the outflow axis, ``bullets" \citep{Bally2016,Zhang+2016}.  The spacing of the bullets indicates that accretion is variable on timescales of hundreds to thousands of years \citep{Bachiller+1991,Lee+2009,Arce+2013}. In the most extreme events, the accretion rate, and hence the source luminosity, rises by several orders of magnitude over a period of years in a brief ``episodic" accretion burst \citep{Audard+2014}.  

Outflows also exhibit precession or changes in direction, providing a window into the angular momentum of accreting material and the impact of binarity on the accretion process \citep{Shepherd+2000,Hirano+2010,Lee+2017}. 
A number of outflows appear to have two components: a highly-collimated component, likely launched close to the protostar, and a wider-angle, slower component that likely arises from the accretion disk \citep{Hirano+2010,Arce+2013}. 

A variety of uncertainties underpin the connection between outflows and accretion \citep{Dunham+2014}. Much of the outflowing material is entrained core material \citep{OffnerChaban2017}, so the outflow is not a direct measure of accreting gas. The typical outflow dynamical time, as measured by the outflow extent and gas velocity, $t_{\rm dyn} \sim L_{\rm out}/(2 v_{\rm out}) \sim 10^3$\,yr, is shorter than the expected protostellar lifetime, and thus provides only a narrow window into the total accretion history \citep{Bally2016}. 
This is probably related to accretion physics (\S\ref{sec:discs}) rather than protostellar dynamics in clusters: Since the velocity dispersions of dense gas and young stars in  star-forming environments are typically of the order 0.1-1~\kms \citep[e.g.,][]{Kirk+2010,Foster+2015}, a protostar would require of the
order of $10^5$ years to move out of a filament or core of 0.1~pc thickness.

Protostellar luminosities provide another constraint on accretion (see Chapter processes). At early times and for low masses, i.e., before the intrinsic luminosity of the protostar becomes significant, the luminosity is directly proportional to the accretion rate. By assuming reasonable properties for the protostellar mass and radii, it is possible to set limits on the accretion rate. However, protostellar evolution remains uncertain in part because it is itself sensitive to the accretion history \citep{PallaStahler1991,Baraffe+2009,Hosokawa+2011}. Observations of clusters of protostars show orders of magnitude scatter, such that on average the luminosity is weakly dependent, at best, on the protostellar class \citep{DunhamPPVI+2014,Fischer+2017}. The difficulty of mapping classes to evolutionary stage further confuses accretion trends 
over time \citep{Robitaille+2006,Dunham+2010,Offner+2012}. 

Time-domain studies of protostellar luminosities are more informative and support the highly variable nature of accretion suggested by outflow observations. Changes in luminosity are observed on timescales of days to decades spanning changes from as little as a few percent to several orders of magnitude in brightness \citep{Rebull+2014,Audard+2014}. Low magnitude, shorter timescale variations, which are quite common, are likely caused by stellar activity or disk occultations, while more extreme and rarer luminosity changes can only be explained by accretion fluctuations \citep{HillenbrandFindeisen+2015}.

Early observations of protostars noted that they were on average about 10 times dimmer than simple accretion models and timescale arguments would suggest \citep{Kenyon+1990,KenyonHartmann1995,Evans+2009}. This became known as the ``protostellar luminosity problem". A variety of theoretical solutions have been proposed that resolve this problem, including episodic and slow accretion (\S\ref{sec:discs}).

Spectroscopic measurements, which become possible once young stellar objects are more than $\simeq 2 \times 10^5$\, yr old, indicate that accretion declines steeply at late times \citep{Hartmann+2016}. However, the accretion rate depends on both age and mass, which are difficult to disentangle due to measurement and model uncertainties. Observations of Balmer continuum, photometry and emission lines suggest $\dot M \propto M_*^\alpha$, where $\alpha = 1.5-3.1$ and $\dot M \propto t^\beta$, where $\beta = -1.6--1.2$ \citep{Hartmann+2016}. 

\subsection{Core-Regulated Accretion Models}

Core-regulated models assume that collapsing gas (compare \S\ref{sec:frag}) efficiently proceeds from $\sim 0.1$\,pc to au scales such that the infall rate is equal to the protostellar accretion rate. Under this assumption, the details of accretion depend only upon the properties of the local gas reservoir. 

\begin{figure}\centering
 \includegraphics[trim=0 4.5in 0 0, width=\textwidth]{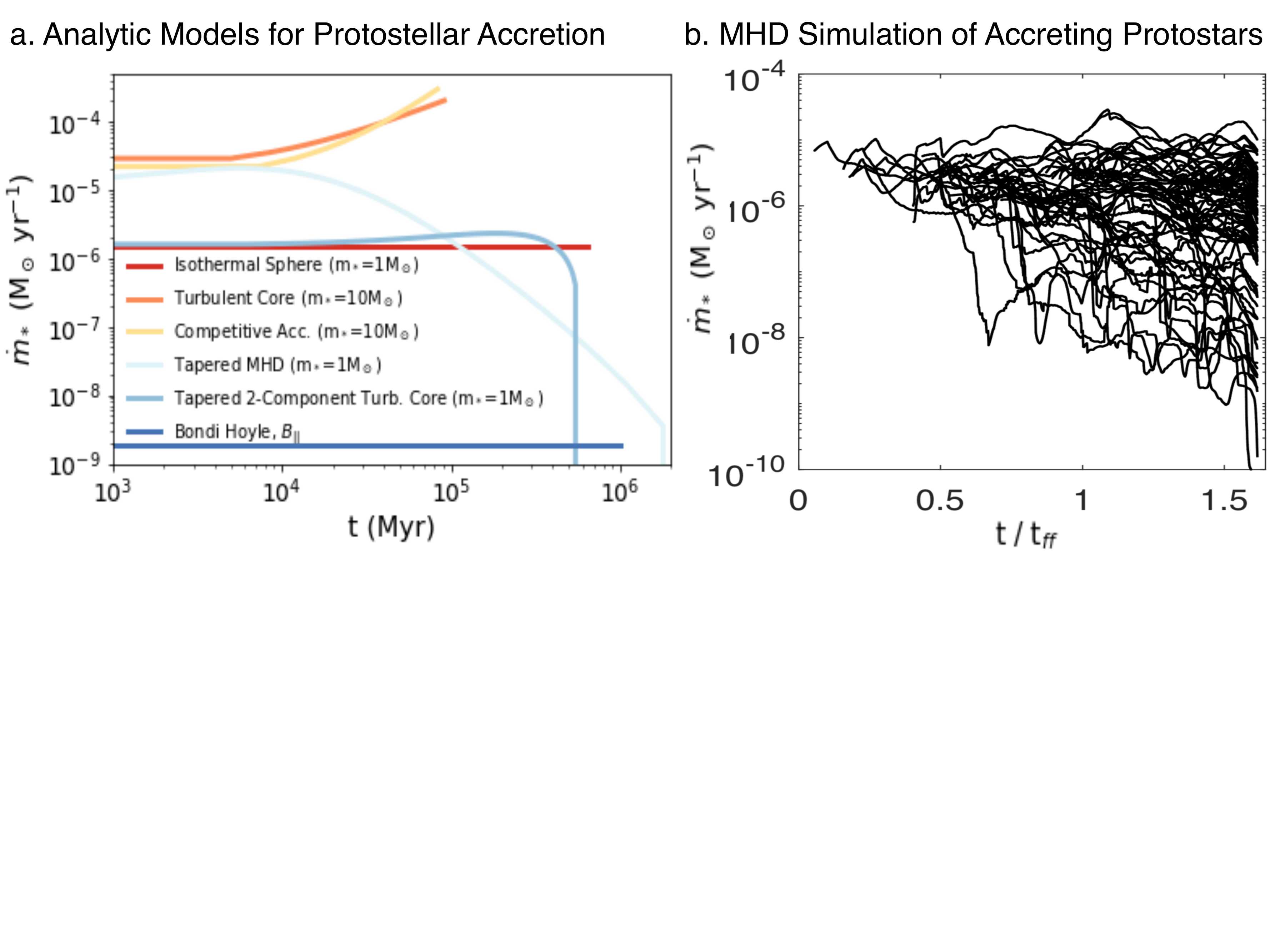}
\caption{ Stellar accretion rates as a function of time. Left (a): different analytic model predictions for protostellar accretion. Right (b): accretion of individual protostars in an MHD simulation of protostars forming in a turbulent giant molecular cloud (figure adapted from \citet{LiPS+18} and reproduced with permission). Turbulence, dynamics and protostellar outflows together significantly modulate the accretion histories.}
\label{fig:acc_models}       
\end{figure}

In the simplest model, collapse is regulated by the interplay of thermal pressure and self-gravity. The accretion rate due to the collapse of a uniform isothermal sphere of gas 
is known eponymously as the Larson-Penston solution \citep{Larson1969,Penston1969}: 
\begin{equation}
    \dot m = 46.9 \frac{c_s^3}{G} = 7.4 \times 10^{-5} \left(\frac{T}{10\,{\rm K}}\right)^{3/2} \msun\,{\rm yr}\e,
\end{equation}
where $c_s$ is the sound speed.
If the gas is isothermal and centrally condensed the infall solution is self-similar and can be written \citep{Shu1977}:
\begin{equation}
\dot m = 0.975 \frac{c_s^3}{G} =  1.5 \times 10^{-6} \left(\frac{T}{10\,{\rm K}}\right)^{3/2}  \msun\,{\rm yr}\e.
\label{eq:iso}
\end{equation}
Figure \ref{fig:acc_models}a shows the isothermal sphere accretion rate and a variety of other analytical predictions as defined below. In these models, the accretion is by nature time-invariant and independent of stellar mass.

These idealised solutions, however, gloss over a great deal of important physics. Cores are observed to be magnetised and turbulent \citep{Crutcher12,Kirk+2017}. They also exhibit velocity gradients indicative of rotation \citep{ChenC+2019,2019ApJ...886..119C}. Moreover, cores are not spheres but are frequently asymmetric, where elongation is likely dictated by the greater filamentary environment that hosts them \citep{Pineda+2010,Arzoumanian+2011}. 
A variety of theoretical models have attempted to address these complications 
\citep{Terebey+1984,Fatuzzo+2004,AdamsShu2007}. The simplest way of modifying equation (\ref{eq:iso}) is to treat turbulence and magnetic fields as contributions to the total pressure support against gravity. For example, replace the thermal sound speed with $c_{\rm eff} = c_s (1 + 2\alpha + \beta)^{1/2}$, where $\alpha =P_{\rm B}/P_{\rm th}$ and $\beta = P_{\rm turb}/P_{\rm th}$ are the ratio of magnetic and turbulent pressure to thermal pressure, respectively \citep{Stahler+1980}. However, both turbulence and magnetic fields are intrinsically anisotropic, which suggests this approach over-simplifies their true impact on the accretion rate and over-estimates their contributions to pressure support.
Also, these models implicitly assume that star formation can be represented by discrete collapsing regions and thus, arguably, are applicable only for isolated, low-mass star formation (cf. \S\ref{sec:frag}). 

The ``turbulent core model" developed by \citet{McKeeTan2003} treats cores as high-column density centrally condensed objects and includes turbulence as an effective pressure. This model predicts that $\dot M \propto M^{1/2}M_f^{1/4}$, where $M$ is the instantaneous stellar mass and $M_f$ is the final mass of the star at the end of accretion. This naturally implies that high-mass stars have higher accretion rates than low-mass stars, form faster and that their accretion {\it increases} in time. Hydrodynamic calculations of high-mass star formation, which adopt high-column density, high-mass cores as initial conditions exhibit these trends \citep{KKMcK12,Rosenea16}. 

Hydrodynamic simulations of forming star clusters paint a very dynamical picture, particularly in clusters with high-stellar densities. In the ``competitive accretion" model protostars begin as small seeds formed by local collapse, which compete with one another for the available gas \citep{Bonnell01}. 
Birth location and dynamical interactions determine the protostellar locations within the gravitational potential well and thus their rate of gas accretion. More massive stars naturally form in the center of the cluster and are best positioned to rapidly accrete gas \citep[Fig.~\ref{fig:age-mass_distribution},][]{Bonnell01,BonnellBate2006}. In this scenario accretion continues until the gas runs out, which occurs on $\sim$ a global free-fall time. As a result, all stars have the same formation time, which is set by the cluster environment. Analytically, this corresponds to accretion rates of $\dot M \propto M^{2/3}$ for stars in gas-dominated potentials \citep{Bonnell01}. At late times when the stellar mass exceeds the gas mass, accretion limits to $\dot M \propto M^{2}$ \citep[compare][]{BallesterosParedes+15, Kuznetsova+17, Kuznetsova+18}. 
Numerical simulations following the formation of massive stars from a 250\,pc scale interstellar medium region suggest that massive stars form over a longer time period via converging, filamentary gas flows \citep{Padoan+2019}. Their cores are less massive than the final stellar mass at any given time, i.e., massive stars do not form from progenitor massive turbulent cores. Inflow rather than competition drives the accretion behavior. 

Stellar accretion is highly variable but does not increase with stellar mass as predicted by both the turbulent core and competitive accretion models. All core-regulated models fall somewhere in the continuum between constant accretion rate and constant accretion time, between the highly dynamical and isolated star formation paradigms. 

\subsection{Disk-Regulated Accretion Models}
\label{sec:discs}

Observations suggest that mass does not pass smoothly from the outer envelope to the protostar but instead accretes in a more variable process as mediated by an accretion disk. Thus, the semi-analytic models outlined above only describe the {\it time-average} accretion behavior. Variability in models that do not explicitly include disk physics
arises purely from variation in the environment and evolution of the host gas reservoir \citep{Padoan+2014,LiPS+18,Padoan+2019}.

The formation of a disk is a direct consequence of angular momentum in the star formation process. In the absence of angular momentum or in the limit of perfectly efficient angular momentum transport accretion disks would not exist. However, observations tell us disks are common \citep{Tychoniec+2018,Andrews+2018}. They act as a repository for high-angular momentum gas
and effectively sort low-angular momentum material, which moves inwards towards the protostar, and high-angular momentum material, which moves outwards.
The two dominant processes for angular momentum transport in disks, viscous torques due to turbulence \citep{BalbusHawley1994} and gravitational instability (GI)  \citep{Toomre1964,LaughlinBodenheimer1994}, both produce variability in the accretion flow. 

Viscous torques require the activation of the magnetorotational instability (MRI), which depends on the local ionization fraction \citep{BalbusHawley1994}. If the gas is not sufficiently ionized then the magnetic field is poorly coupled, reducing the efficacy or shutting off the MRI entirely \citep{BlaesBalbus1994}. Thus, MRI-regulated accretion disks may undergo periods with little or no accretion during which material builds up in the disk, followed by periods when the gas is thermally ionized initiating a burst of accretion \citep{Zhu+2009b}. During these bursts accretion may be elevated by several orders of magnitude, similar to observed FU Ori bursts \citep{Audard+2014}.

The requisite ionization for the MRI may be provided by the parent star and its environment
including FUV radiation, x-rays, and cosmic rays \citep{UmebayashiNakano1981,Semenov+2004,Glassgold+2007,PerezChiang2011}. Consequently, disk surface layers are generally strongly ionized such that accretion continues in a layered fashion, where gas accretes in the surface layers  while the disk mid-plane remains predominantly neutral and is an MRI ``dead zone" \citep{Gammie1996}. High periods of accretion may in turn increase the x-ray and cosmic-ray ionization towards the disk mid-plane, prompting accretion deeper in the disk and boosting the magnitude of the accretion burst \citep{Offner+2019}.

Gravitational torques, which are the dominant transport mechanism in the outer disk, may also prompt large accretion variations \citep{KratterLodato2016}. If mass builds up in the inner disk, the disk may undergo GI and form small clumps. If these clumps migrate inwards they produce burst events as they accrete onto the star \citep{VorobyovBasu2005,VorobyovBasu2006}. Mild GI, in the form of spiral arms, to severe GI, which causes catastrophic disk fragmentation, produce accretion variability from factors of a few to orders of magnitude \citep{Audard+2014}.

Dynamical interactions between stars or close binary companions can also produce accretion variability \citep{AdamsLin1993}. Close passage gravitationally perturbs the disk, prompting instability and elevated accretion \citep{Bonnell1994}. Finally, variation of the angular momentum of the infalling gas on larger scales may also create luminosity variations, either through direct accretion \citep{Padoan+2014} or by affecting disk properties \citep{LeeJ+2017}. 

The frequency and magnitude of disk-mediated accretion bursts depend both on disk microphysics and the larger disk environment \citep{Kratter+2010a}. Current observations show a heterogeneous distribution of large, small, smooth and structured disks. Likely both, GI and MRI, play a role in disk evolution \citep{Armitage+2001,Zhu+2009}.
The corresponding scatter in protostellar luminosities provides one solution for the protostellar luminosity problem \citep{Kenyon+1990,OffnerMcKee2011,DunhamVorobyov2012,Padoan+2014}.

\subsection{Feedback-Regulated Accretion Models} \label{sec:feedbackregulated}

Stellar feedback, in the form of protostellar outflows, winds and radiation, also shapes the accretion process, either by reducing the mass reservoir available for accretion \citep[as found in observational and numerical work, see e.g.][]{Dalea15b,GinsburgEtAl2016} or by dispersing bound gas and halting accretion altogether. The earliest semi-analytic model for feedback-regulated accretion weighed the competition between accretion and outflow feedback \citep{NormanSilk1980}. Feedback-regulated models are often formulated more generally in terms of a distribution of stopping times or probabilities, which has the advantage that the model can be agnostic about the particular mechanism halting accretion. For example, 
several more recent models 
assume accretion durations follow the probability distribution, $f(t) = 1/\tau e^{-t/\tau}$, where $\tau$ is the mean accretion time (of the order of $10^5$~yr).  Such models can reproduce the stellar IMF and match the observed protostellar luminosity distributions without appealing to overly long accretion times or significant periods of episodic accretion \citep{BasuJones2004,Myers2009,Myers2012}.

A variety of hydrodynamic simulations of accreting protostars including protostellar outflows have been carried out, which demonstrate that outflows can indeed efficiently expel 30-60\% of the dense core material and reduce overall star formation efficiencies by $\sim 30$\% \citep{Hansen+2012,MachidaHosokawa2013,OffnerArce2014,Federrath2015,OffnerChaban2017,Tanaka+2017}. 
Simulations of isolated dense cores including protostellar outflows find that the main phase of accretion continues for 0.3-0.5\,Myr, depending on the degree of turbulence and magnetic field strength 
\citep{MachidaHosokawa2013,OffnerArce2014,OffnerChaban2017}. The accretion rate of a protostar accreting within a turbulent, magnetised dense core can be described in terms of the current protostellar mass, $m$, and its final mass, $m_f$:
$\dot m = m_0 \left( \frac{m}{m_{f}} \right)^{1/2} m_f^{3/4} \left[ 1- \left( \frac{m}{m_f} \right)^{1/2}\right]^2,$
where $m_0 \propto \Sigma_c^{3/4}$ is a constant coefficient related to the surface density of the core, $\Sigma_c$, and both $m$ and $m_f$ are in solar masses \citep{OffnerChaban2017}. This is effectively the predicted turbulent core model accretion rate \citep{McKeeTan2003}, tapered by a multiplicative factor. While the final masses are influenced by the core magnetic field and turbulence, the accretion history can be analytically described independently of the gas physical properties.
Simulations of the impact of outflows on accretion within forming star clusters find wide variation in the accretion histories as shown for example in Figure \ref{fig:acc_models}b with some accretion rates steadily declining over time to $\dot m = 10^{-8}\,\msun$\,yr\e and others declining and then rising again due to protostellar dynamics \citep{LiPS+18}.

Feedback, turbulence and gravitational interactions
may all
play important roles in setting the accretion histories of individual stars. 
These same processes also drive the global evolution of the molecular cloud, gas dispersal (\S\ref{sec:gd}) and star cluster dynamics (\S\ref{sec:nbody}). Thus, it is not possible to separate the formation of individual stars from the structure and evolution of the larger star cluster, i.e., whether it is a strongly bound cluster or a quickly dispersing association.

However,
once the initial formation of the stars is
completed, stars and gas effectively
de-couple. In the following two sections
we therefore first review studies that
focus on the evolution of the gas,
and then ones that treat the dynamics
of the stars. The limitations of these
approaches will become obvious when
the transition from the star formation
epoch will be considered and contact
to models that specifically target the transition phase will be made.

\section{Feedback and gas dynamics}\label{sec:gd}

After the initial star formation process, clusters become exposed, i.e., no dense gas is found in clusters from this stage onward. The process of a cluster becoming exposed
may be driven by collective stellar feedback and may influence the dynamics of the stars. Later, star cluster winds can convey feedback energy to larger scales. Cooling flows have been discussed in the context of secondary star formation episodes, although age spreads in clusters are small, such that secondary star formation is likely restricted to associations.

Star clusters have a closed tidal surface and usually contain a focal 
point, the minimum of the gravitational potential. It is therefore
generally expected that a global pattern for the gas dynamics 
will form, which may in principle be inflow, outflow, or 
hydrostatic equilibrium. Contrary to galaxies, 
(even approximate) hydrostatic equilibrium is probably not relevant for star clusters.

\subsection{Impossibility of hydrostatic equilibrium}
To see this, we present a simple argument and show that starting from a
situation close to hydrostatic equilibrium, 
stellar feedback would alter the gas properties quickly. Either cooling would take over leading to inflow, or heating, leading to outflow.
Let us start with the hydrostatic equilibrium condition \footnote{$\Phi$: gravitational potential, $r$: radius, $\rho$: gas density, $p$: pressure}:
\begin{equation}\label{eq:hseq}
    \frac{{\rm d}\Phi}{{\rm d}r} = - \frac{1}{\rho}\frac{{\rm d}p}{{\rm d}r}
\end{equation}
Approximating gradients by the absolute change out to the half-mass radius,
we can write eq.~(\ref{eq:hseq}) 
as\footnote{$M$: cluster mass, $\mu$: mean molecular weight}: 
$G(M/2)/r_\mathrm{h}=p/\rho=k_\mathrm{B}T/(\mu m_\mathrm{p})$.
Radiative cooling will reduce the gas pressure.
To maintain hydrostatic equilibrium, the 
cooling time\footnote{$k_\mathrm{B}$: Boltzmann constant, T: temperature, n: particle density, $\Lambda$: cooling function.}
$t_\mathrm{c}=k_\mathrm{B}T / (n \Lambda)$
therefore must at least exceed the crossing time\footnote{$r_\mathrm{h}$: half-mass radius, $\sigma$: line-of sight velocity dispersion.} $t_\mathrm{x} = 2 r_\mathrm{h}/\sigma$ \citep{KHS19}.
Using also the definition 
\begin{equation}\label{eq:sigma}
\sigma^2 = GM/(\eta r_\mathrm{h})
\end{equation}
with $\eta = 7.5$ for a \citet{1911MNRAS..71..460P} model, we arrive at the constraint
\begin{eqnarray}
    n &<& \frac{G^{3/2}\mu m_\mathrm{p}M^{3/2}}{\eta^{1/2}\Lambda r_\mathrm{h}^{5/2}} \nonumber\\ 
    &=& 65 \, \mathrm{cm}^{-3}
    \left(\frac{\Lambda}{10^{-27}\,\mathrm{erg}\,\mathrm{cm}^3\mathrm{s}^{-1}}\right)^{-1}
    \left(\frac{M}{10^5M_\odot}\right)^{3/2}
    \left(\frac{r_\mathrm{h}}{3 \, \mathrm{pc}}\right)^{-5/2}\, .
    \label{eq:nlim}
\end{eqnarray}
For the relevant densities, $\Lambda$ is of the order of 
$10^{-27}\,\mathrm{erg}\,\mathrm{cm}^3\mathrm{s}^{-1}$ \citep{BiSt19},
which we have used for the scaling in eq.~(\ref{eq:nlim}).

The immediate effect of stellar feedback is to add 
mass and energy to the intracluster gas. Hydrostatic equilibrium may only be maintained, if the energy input matches the energy loss via gas cooling. The particle density in the cluster increases at a rate \citep[e.g.,][]{MB03}:
\begin{eqnarray}
    \dot{n}=\frac{\alpha M/2}{4\pi r_\mathrm{h}^3/3} 
    =
    56 \,\mathrm{cm}^{-3}\mathrm{Myr}^{-1}
    \left(\frac{\alpha}{10^{-16}\,\mathrm{s}^{-1}}\right)
    \left(\frac{M}{10^5M_\odot}\right)
    \left(\frac{r_\mathrm{h}}{3 \, \mathrm{pc}}\right)^{-3}\, .
    \label{eq:ndot}
\end{eqnarray}
Here, we have scaled the mass loss factor $\alpha={\dot M}/M$ to 
$10^{-16}\,\mathrm{s}^{-1}$, a value that would be expected
for a very young ($\approx$~Myr) stellar population
\citep{Leithea99,GCK05,Krausea13b}.

Therefore, within a short timescale compared to the timescale
of a cluster's evolution, stellar feedback would increase the 
gas density beyond the cooling limit. Hydrostatic equilibrium could
then only be maintained, if the energy input was spatially fine-tuned 
and arranged to increase in time as required for the increasing
cooling rates. Since cooling rates are determined by atomic physics
and energy input by stellar physics, this will not be the case.

The late time evolution of $\alpha$ can be approximated as
$\alpha=4.7\times10^{-20}\,\mathrm{s}^{-1}$ 
$(t/13~\mathrm{Gyr})^{-1.3}$
\citep{MB03}. Therefore, if at late times the cluster was for some reason
in a state of hydrostatic equilibrium, it would take longer
for the stellar feedback to increase the gas density beyond the
stability limit. However, the relevant timescales also grow, such
that the cluster would always become unstable on a timescale that
is shorter than its age.
The analysis depends only weakly on cluster radius. For smaller masses, 
the argument becomes stronger.
Therefore, we can conclude that stellar feedback generally inhibits
hydrostatic equilibrium in star clusters at all times.

\subsection{Initial gas clearance}\label{sec:gasclear}
How much gas is left over from the star formation process and how violently this
is removed from the cluster has wide-ranging implications for star formation.
From abundances and ages of clusters and associations, \citet{LL03} concluded 
that stars generally form in dense clusters and get dispersed due to violent 
gas expulsion and the associated change in gravitational potential
({\em infant mortality}).
Subsequent work has superseded this initial picture, showing that the statistics depend crucially on the 
surface density threshold for the definition of star clusters 
\citep{Bressea10}, as well as the initial gas density at which stars are forming \citep{Kruijssen12}. Many recent studies show that star formation
proceeds at a variety of densities and spatial scales 
\citep[e.g.,][]{Bastea07,Sunea18,Rodrigea19} and detailed analysis of OB 
associations shows that they did not evolve from significantly smaller
structures \citep{Wrightea14,WardKruijssen2018,Wardea19}.
A good example is the {\it Gaia} study of the closest OB~association,
Sco-Cen OB2 \citep{WM18}, for which alternative formation scenarios based on multi-wavelength
observations have been suggested \citep{Krausea18b}.
While the formation of bound clusters and their dispersal may be less common than once thought
\citep[compare also][]{Kruijssen12, Krumhea19}, it is still interesting to ask what fate
the gas experiences and what roles it can play in any given cluster.

\subsubsection{Gas expulsion}\label{sec:gex}
We use the term {\em gas expulsion} to refer to a special kind of gas removal,
where a significant mass of gas ($\gtrsim 50\%$ of the total mass)
is removed quickly (compared to the crossing time for stars, i.e.\ impulsively) from
the cluster such that some or all stars are left unbound and escape
\citep{Hills1980}.
The only situation where this can
happen is at the end of the initial formation
of the star cluster from the primordial gas cloud, 
hence the 
frequent use of the term primordial gas expulsion. 

Assuming that the gas retains the same spatial profile as the young stellar cluster, the effect of primordial gas expulsion has been studied extensively
in pure $N$-body simulations, where the stars are represented by a large
number of gravitationally interacting bodies and the gas by a smooth
potential that is varied in time 
\citep[e.g.,][for reviews]{PZMcMG10,Banerjea17}. 
\citet{BK07} show in a large parameter study 
that most clusters are completely destroyed or lose a 
substantial number of stars. Those that survive have expanded 
by a typical factor of 3-4. More recent N-body simulations vary, e.g.,
the kinematic state at gas expulsion or the level of substructure
\citep{SmithRea13,Fariea15} and find that cluster dispersal becomes more 
difficult in more realistic scenarios \citep{Fariea18}.

\begin{figure}\centering
  \includegraphics[width=0.98\textwidth]{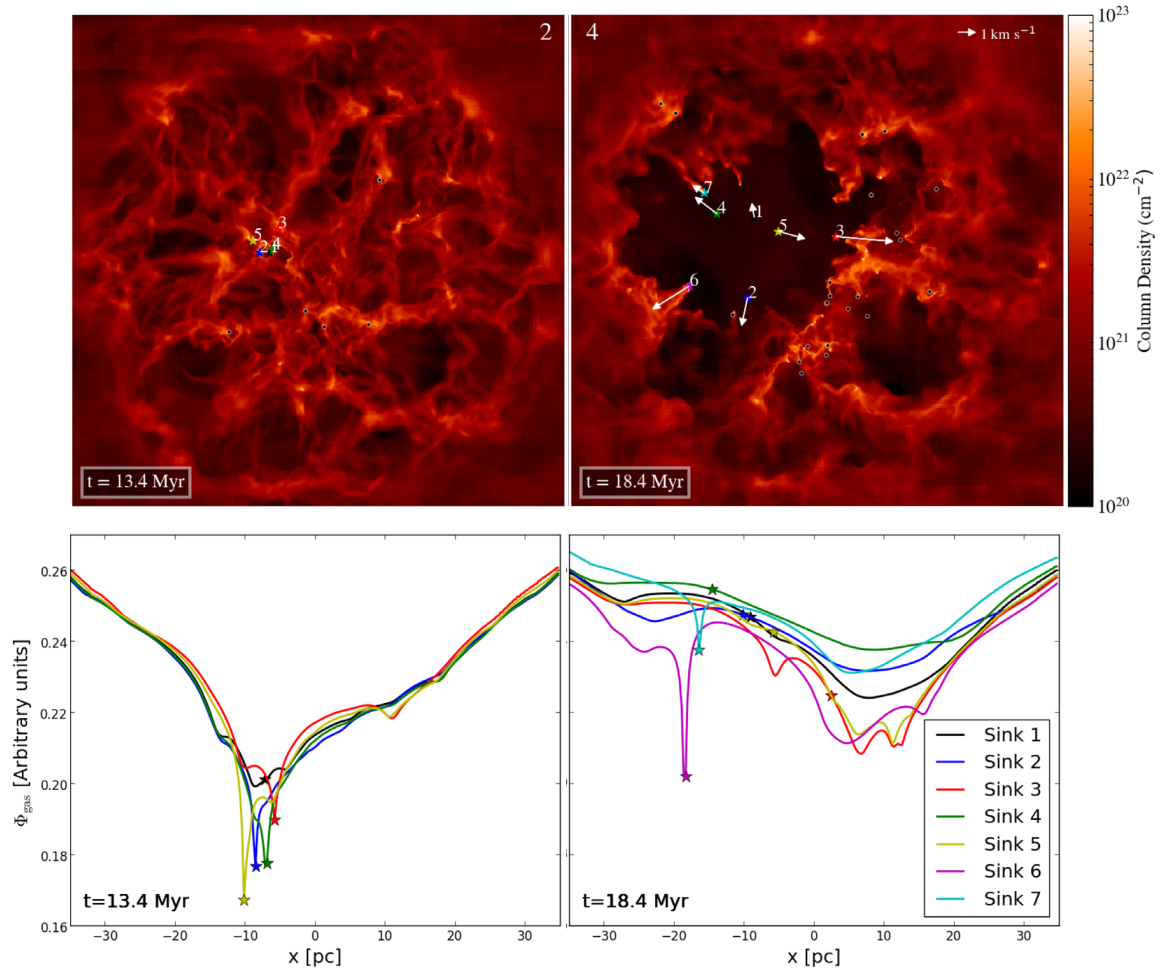}
\caption{Two time frames of a numerical simulation of the evolution of a molecular cloud with stellar feedback. Upper panels are maps of the column density. Lower panels are the corresponding $x-$profiles of the gravitational potential at the $y$ position of each sink particle. As the expansion of the \HII\ region proceeds, the gravitational potential is flipped-up, and thus, the stars are pulled out toward the edges. Figure from \citet{ZamoraAvilez+19}.
}
\label{fig:gf-sim}       
\end{figure}

Hydrodynamic simulations by \citet[][]{Geen+18} and \citet[Fig.~\ref{fig:gf-sim}]{ZamoraAvilez+19} have shown that the dispersal of the parental molecular cloud could have a ``gravitational feedback'' effect on the newborn stellar cluster: feedback from the newborn massive stars expels the gas from the collapse centre. Since neither the parental clouds, nor the formed shells are distributed symmetrically around the \HII\ region, net forces can even accelerate the stars towards the edges of the cavity and may produce a ``Hubble flow-like'' ($v\propto r$) expansion. 

\subsubsection{Observed kinematics in young star clusters}
\label{sec:obskinYSCs}
Several candidates for stellar groups undergoing expansion or dispersal related to gas expulsion
have been found with {\it Gaia} Data Release 2,
for clusters with masses up to $10^4 M_\odot$: 
 \citet{Kuhnea19} study the kinematics
of 28 young stellar groups with typically 100 stars with proper motion
measurements each. For 75\% of their objects, they find
a positive offset of the generally Gaussian distributions of the 
cluster-centric radial velocities, i.e.\ an expansion of the system.
Some of their groups are likely unbound 
and may have formed as associations
\citep[compare also][]{Braviea18,Wrightea19}, while some could have
undergone an expansion phase and are settling in virial equilibrium.
There is an interesting variety in expansion states also for compact
systems. 
\citet{Kuhnea19} find the partially embedded Orion Nebula cluster (ONC)
($r_\mathrm{h}=0.9$~pc) to be only slowly expanding 
($v_\mathrm{out}=0.43\pm0.20$~km~s$^{-1}$), but  the likewise partially
embedded cluster Cep B ($r_\mathrm{h}=1.4$~pc) to 
be expanding at more than 
twice this rate and to clearly show a Hubble-law-like behavior (compare \S\ref{sec:gex})
as expected to develop in dispersing star clusters.
\citet{Karnea19} provide detailed kinematics for two expanding sub-clusters of
Cep~OB3b, arguing that the clusters might lose 25-65\% of their stars, before
re-settling in virial equilibrium.

Young ($\approx 10$~Myr), exposed, massive ($\gtrsim 10^4 M_\odot$) star clusters also appear frequently to 
have velocity dispersions above the expectation for virial equilibrium, 
given the mass expected for the observed luminosity and age 
\citep[e.g.,][]{BG06,GB06,Gielea10,PZMcMG10}. This has been discussed as evidence 
for dissolution after gas expulsion \citep{GB06}. However, $N$-body simulations show that many of these clusters would have re-virialised by the time of observation 
\citep{BK07,Gielea10,PZMcMG10}. An interpretation in terms of a large contribution 
from binaries to the velocity dispersion \citep[compare,e.g.,][]{Leighea15,OKP15} 
seems more plausible \citep{Gielea10,Cottea12,HenBruea12}. 

\subsubsection{Gas expulsion in massive star clusters}\label{sec:gasexpul}
It can be shown that there exist a critical 
compactness $M/\rh$ above which gas expulsion
with associated dispersal of stars can no longer work in a star cluster even if the gas dominates the gravitational potential at the time when massive star feedback becomes effective: while the gravitational binding energy $E_\mathrm{b}$
is proportional 
to\footnote{$\epsilon_\mathrm{SF}$: stellar mass $M$ over total mass (stars + gas) of an embedded cluster} 
$(1-\epsilon_\mathrm{SF}) M^2/r_\mathrm{h}$, the cumulative feedback energy by winds
and supernovae at any given cluster age is only linear in the mass: 
$E_\mathrm{f}\propto \epsilon_\mathrm{SF}M$.
Therefore, gravity must eventually win. 

If we demand that for successful gas expulsion to happen,
the provided feedback energy must exceed a critical energy 
proportional to the binding energy, i.e.,
\begin{equation}
    E_\mathrm{f} > a^{-1} E_\mathrm{b} \, ,
\end{equation}
with a constant $a^{-1}$ that will depend on the details of the feedback physics,
then we can derive a critical star formation efficiency, defined
here as the ratio of stellar mass in the cluster to its total mass during
the embedded phase, for gas expulsion to succeed:
\begin{equation}
    \epsilon > \epsilon_\mathrm{crit}(C_5) = a C_5 
    \left(-\frac{1}{2} + \sqrt{\frac{1}{4}+\frac{1}{aC_5} \, ,}\right)
\end{equation}
where we have defined the compactness index as
\begin{equation}
    C_5 = \frac{M/r_\mathrm{h}}{10^5M_\odot\,\mathrm{pc}^{-1}}
    = \left( \frac{\sigma}{7.5 \,\mathrm{km\, s^{-1}}}\right)^2
\end{equation}
and used eq.~(\ref{eq:sigma}) in the final equality above. 

\begin{figure}\centering
  \includegraphics[width=0.45\textwidth]{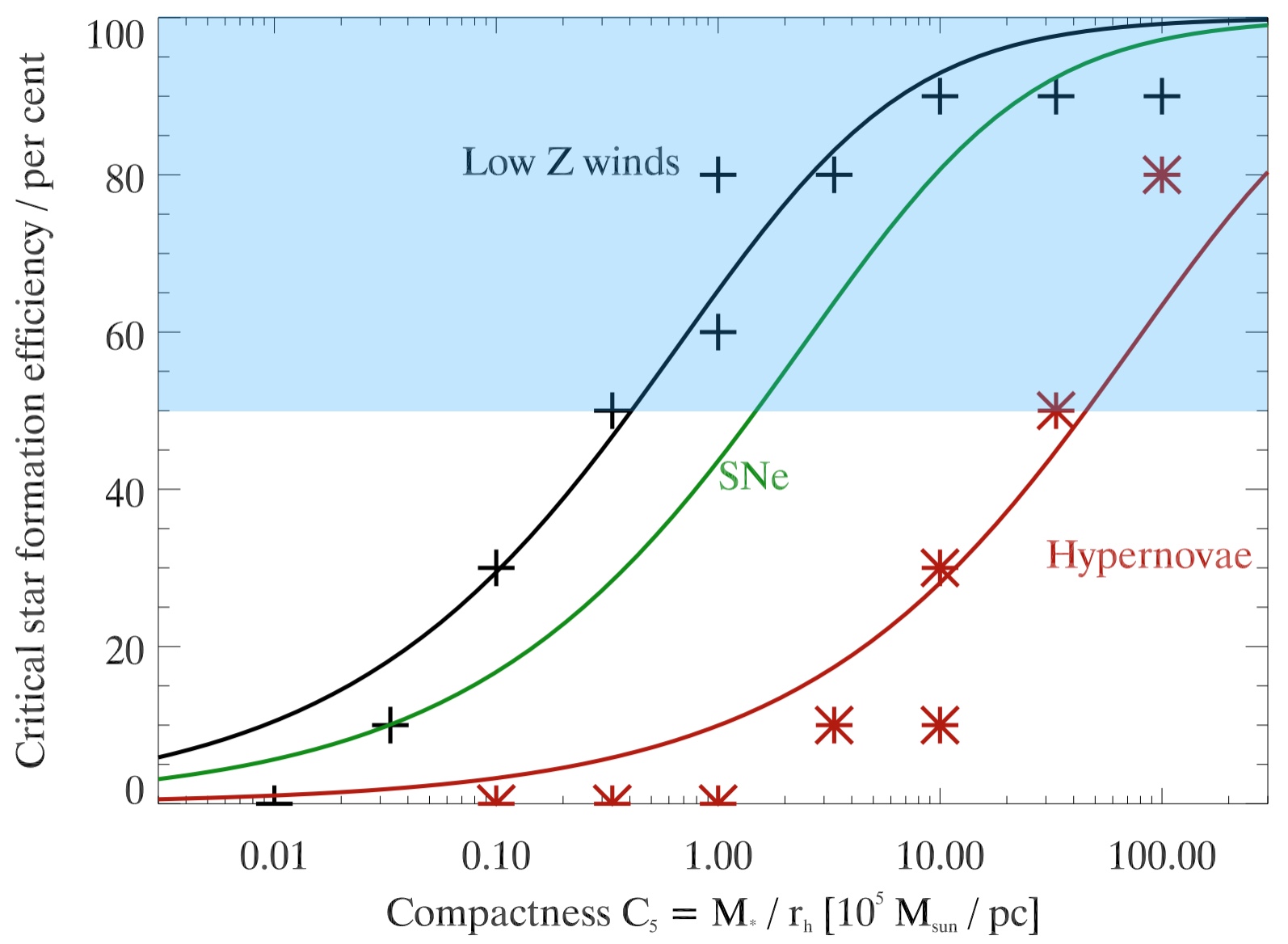}
\caption{Gas dynamical constraint on the star formation efficiency for successful gas expulsion. If the star formation efficiency, i.e. the ratio between stellar and total mass in the embedded cluster, is less than the given value, the given type of
feedback will not be able
to expel the gas on the crossing timescale of the cluster. N-body simulations
find an upper limit of the star formation efficiency of 50\%, if gas expulsion
is to significantly affect the stars (non-shaded region). 
This results in an upper limit on the compactness 
of a star cluster, $C_5$, above which gas expulsion cannot lead to significant
expansion, loss of stars or dispersal. Curves for three assumptions 
on the type 
of feedback responsible for the gas expulsion are shown: Stellar winds at
metallicity $\feh=-1.5$ (black), supernovae (green) and 
a burst of hypernovae (red), as an extreme upper limit for gas expulsion via
stellar feedback.
Based on Fig.~4 in \citet{Krausea16a}.}
\label{fig:epsSF_crit}       
\end{figure}

The function $\epsilon_\mathrm{crit}(C_5)$ tends towards zero for small $C_5$ ($\sigma^2$)
and towards one for very high cluster compactness. \citet{Krausea16a} have shown
that a thin-shell superbubble model reproduces this equation (Fig.~\ref{fig:epsSF_crit}).

Thin-shell superbubble models \citep{Krausea12a,Krausea16a} compute the
kinematics of the supershell assuming some prescription for the energy input
and spherical symmetry. They can, however, take 3D effects into account by
evaluating a criterion for the shell's acceleration. The shell will be 
destroyed by the Rayleigh-Taylor instability as soon as modes comparable to the
size of the shell become unstable. The hot, pressurised bubble interior then 
escapes through holes in the shell, and the dense shell gas falls back.
The more stars there are compared to the amount of gas, the stronger the
feedback, and the easier to push out the gas without making the shell unstable.

Successful gas expulsion therefore requires the star formation efficiency to
be above a certain limit. There is, however, also an upper limit ($\approx 50\%$),
if one wants gas expulsion to affect the stellar population. Both constraints
together imply that only clusters with a compactness below a critical value
can suffer expansion or even dispersal due to gas expulsion. For both,
solar metallicity winds and supernovae, \citet{Krausea16a} show that the critical
compactness index is $C_5\approx 1$ ($\sigma = 7.5$~km~s$^{-1}$, 
also compare Fig.~\ref{fig:epsSF_crit}). 

The thin shell models effectively correspond to the assumption of 
maximum efficiency for stellar feedback: the hot gas is assumed to be 
always more central than the cold gas, thus maximising the outward
push on the cold gas. The spherical shell prevents hot gas from
escaping, thus all of it can be used to act on the cold gas.
Finally, \citet{Krausea12a,Krausea16a} assume 80\% of the released feedback
energy to be radiated away, thus 20\% to be available for gas dynamics.
This is likely a generous assumption, given the high efficiency of mixing
and associated radiative losses seen in recent 3D superbubble simulations with
time-dependent driving \citep{Krausea13a,Vasilievea17,Gentrea19}.

Stellar winds become less efficient at low metallicities $Z$, 
their energy output scaling with $Z^{0.7}$ \citep{MM12}.
If stellar winds (augmented by photoionisation and radiation pressure effects further away 
from the massive stars, compare below) dominate feedback in young clusters,
rather than supernovae,
for which there is some evidence
from the timescales observed for massive clusters to become exposed
\citep{Hollyhea15,Sokalea16,Kruijea19a,Chevea19}, the critical compactness index
becomes smaller at low metallicities, $C_5=0.3$ at $\feh=-1.5$.
It is also possible to increase it by extreme assumptions on stellar feedback.
If the most massive stars in a cluster exploded as hypernovae, 
all releasing ten times the conventional supernova energy output of $10^{51}$~erg 
\citep[e.g.,][]{Mazea14,LueHJea18}, this would increase the critical
compactness index to $C_5\approx 30$. 
A more comprehensive analytic treatment by \citet{MJ15} 
that takes into account accretion and various feedback processes
separately find the threshold at 3~km~s$^{-1}$ ($C_5=0.2$).

\subsubsection{Slow gas clearance}

\citet{Crockea18} consider the effect of the radiation pressure taking into account
re-radiated infrared radiation due to the presence of dust. They argue that indirect
radiation pressure on dust would first expand the gas gently, and that direct radiation
pressure would later, but still before the first supernova, expel the gas on the 
dynamical timescale. For favourable assumptions, they find
a maximum stellar surface density of $10^4M_\odot$~pc$^{-2}$ at which 
up to $\epsilon_\mathrm{SF}=50\%$ of the mass in a cluster can be stars without them forcing the remaining 
gas mass out of the cluster. Taking a typical cluster radius of 1 pc converts this result 
to a compactness index $C_5 = 0.3$. Hence, radiation pressure is expected to be 
somewhat less effective at expelling gas than stellar winds, but of comparable
order of magnitude \citep[compare also][]{Reisslea18}.

\citet{Rahnea17} use a self-gravitating thin-shell model to predict gas removal
in clusters with $0.05<C_5<100$. They find comparable contributions
from radiation pressure, winds and supernovae, with radiation pressure dominating
at the high-mass end
\citep[compare also][]{KKO16}. 
\citet{Rahnea19} use an updated treatment of the hot gas pressure,
similar to \citet{Krausea12a,Krausea16a}. They also investigate the effect of the power-law
slope of the initial gas distribution outside the core radius.
For $0.025<C_5<2.5$ they find minimum star formation efficiencies for gas removal
in the percent range, with the exception of their most concentrated clouds with
their steepest gas density power-law index of -2, where it can reach 50\%.
As they consider only gas removal on any timescale, and not the specific condition
for gas expulsion on the dynamical timescale, their critical star formation efficiencies
are somewhat lower than the ones of \citet{Krausea12a,Krausea16a} despite them including
radiation pressure into their calculations. 

\subsubsection{Gas clearance in 3D hydrodynamics simulations}
Multi-dimensional simulations of star cluster formation that take into 
account the actual formation of the stars and follow feedback from 
individual massive stars typically find a reduced effect of feedback compared
to the much more idealised works above. \citet{Dalea15b} study the evolution
of a turbulent molecular cloud with photoionisation and conservatively
implemented stellar wind feedback using smooth particle hydrodynamics. 
They report a variety of conditions for star 
formation, including tenuous and very dense regions, with
the overall number of expelled stars remaining low.
\citet{Gavagnea17} conducted a similar study using adaptive mesh refinement
hydrodynamics together with photoionisation from individual stars in an initially subvirial cloud.
They report runs with different feedback strength. The fraction of 
unbound stars depends only weakly on the feedback strength, and ejections
are mainly due to gravitational star-star interactions. Surprisingly,
their star cluster without feedback disperses at the end of the simulation,
whereas the cluster with the strongest feedback forms a subvirial system,
despite 80\% of the gas being ejected. This is, because the feedback
efficiently slows the overall collapse, such that the stellar density remains lower,
and less dynamical interactions between stars take place.

The accuracy with which the strength of feedback is predicted by these models may be subject to further improvement. The different 
feedback processes (accretion radiation, protostellar jets and outflows, photoionisation,
radiation pressure, stellar winds and supernovae) require very different computational methods.
The simulations discussed above all include photoionisation. \citet{Dalea15b} exclusively use the momentum from stellar winds. This is an underestimate, because the energy in the winds will not be entirely radiated away, but produce some
additional momentum. That the simulations generally underestimate feedback is underlined
by the fact that many runs do not terminate star formation \citep[e.g.][]{Dale17} within an observationally required 
time frame of 3 Myr \citep{Chevea19}. This is particularly relevant, given that
the timescale of gas loss strongly affects any expansion or dispersal 
\citep{SmithRea13}. The virial state is expected to have a strong
influence on the fraction of bound stars \citep{Fariea15}. Hence, simulations
with subvirial clouds, only, \citep{Gavagnea17} cannot provide the full picture.

More recently, \citet{LiHea19} simulated star cluster formation from turbulent clouds
in different kinematic states with a moving mesh hydrodynamics code. 
In each run, they form a variety of 
stellar structures, hierarchically merging into bigger ones.
They apply feedback via mass and momentum deposition
around each star, which is varied within a factor of 20. The latter range reflects the still
existing uncertainty on the feedback strength. Gas expulsion with associated dispersal of stars
seems to occur in some of their simulations with the highest level of feedback.
Most of their simulations do, however, not show a strong unbinding of stars due to bound structures
being generally subvirial prior to the gas expulsion treatment.

\subsubsection{Gas clearance summary}
Analytical and semi-analytical models combined with N-body simulations
tend to overestimate the 
effects of stellar feedback and predict strong gas 
expulsion effects with cluster expansion and dispersal for 
a virial velocity dispersion $\sigma<3-7$~km~s$^{-1}$. They
firmly exclude strong effects of gas expulsion on the cluster stars
for $\sigma>7$~km~s$^{-1}$ unless one assumes non-standard
mechanisms. Multi-dimensional simulations tend to underestimate
feedback and usually see little effects of gas expulsion and a small
amount of unbound stars. However, tuning up the feedback strength,
such effects have also been reported \citep{ZamoraAvilez+19,LiHea19}.
{\it Gaia} stellar kinematics observations suggest that gas 
expulsion may be responsible 
for expansion and possibly dispersal of some stellar groups, while compact clusters
appear unaffected, implying slow gas outflow or exhaustion of gas turned into newly formed stars. All reported cases where expansion or dispersal
may take place have $\sigma<3$~km~s$^{-1}$, consistent with 
the theoretical constraints.

These results are in good agreement with direct
observations of gas and stars in young massive clusters and progenitor clouds: there are no clouds compact enough, so that a young massive cluster could form with the same structure as that cloud
\citep{Longmea14,Walkerea15,Walkerea16}.
The implication is that star formation has to proceed as the cloud collapses (compare \S\ref{sec:frag}), which has been termed the `conveyor belt' model of cluster formation \citep{Longmea14}. As gas clouds fall in, they can already be forming stars, with a star formation efficiency that peaks within the regions of the highest densities. This leads to local gas exhaustion in these regions and limits the effects of gas expulsion on the virial state of the resulting cluster, producing high bound fractions and thus compact cluster formation \citep{Kruijssen12}. Low central gas fractions in an embedded cluster are indeed reported by \citet{GinsburgEtAl2016}. Gas exhaustion has also been seen in cluster formation simulations
\citep{GirichidisEtAl2012a,Kruijea12a,Dalea15b}.

\subsection{Steady-state cluster winds}\label{sec:sswinds}
Star cluster winds form, where the gas in the
cluster is heated faster than it can cool. This is usually
expected for the epoch just after the gas has
been cleared from the cluster, either because most of it has
been accreted on to the stars (exhaustion) or, 
because the feedback processes removed it from the cluster.

If there is a significant number of massive stars in the 
cluster, all driving winds and exploding as supernova
in more or less regular intervals, one can assume that
the energy is efficiently thermalised in the local interactions, 
and the mass input from the various wind sources to be
smoothed out over the size of the cluster.
\citet{CC85} developed a classical steady-state wind model 
that applies to this situation.
Important assumptions in the model are:
\begin{enumerate}
    \item Spherical symmetry.
    \item The region of interest comprises a large number
    of stars (sources of mass and energy), such that individual
    sources interact locally and we can describe the gas physics
    using smooth mass and energy input functions $q(r)$ and $Q(r)$,
    respectively.
    \item Top-hat flat source profile, 
    i.e.\footnote{$\dot{M}$: total mass loss rate; 
    $\dot{E}$ total energy release rate; $V=4\pi r^3/3$.}, 
    $q(r)=\dot{M}/V$,
    $Q(r)=\dot{E}/V$ for $r<R$, and $Q(r)=q(r)=0$, otherwise.
    \item Gravity is negligible.
\end{enumerate}

Assumption~2 above restricts the theory effectively to 
massive star clusters (and galaxies, of course). For the
wind phase, this is because of the strong dependence of
the stellar wind strength on the stellar mass. For example,
\citet{Krausea13a} show in 3D hydrodynamics simulations
that for a group that harbours star of 25, 30 and $60 M_\odot$
\citep[typical for a $1000 M_\odot$ cluster using the initial mass function from][]{Kroupea13},
the $60 M_\odot$ star completely dominates the gas dynamics
as long as it exists. Interacting stellar winds and supernovae
will heat the cluster to typically, $10^7$~K, which corresponds 
to a sound speed of \revI{$\approx 500$}~\kms. For a typical cluster diameter of,
say, 10~pc, the dynamical timescale is then \revI{20,000}~yr.
If we require one supernova per dynamical timescale,
we need roughly \revI{$1500$} massive stars, which we expect
for a cluster with \revI{$\approx 10^5 M_\odot$}. Steady-state winds
in clusters are therefore frequently referred to as 
super star cluster winds. \citet{CRR00} show using 3D
hydrodynamics simulations that in a cluster with 30 massive
stars with similar properties, the 1D case with smooth source
functions is approximately recovered.

\begin{figure}\centering
  \includegraphics[width=0.45\textwidth]{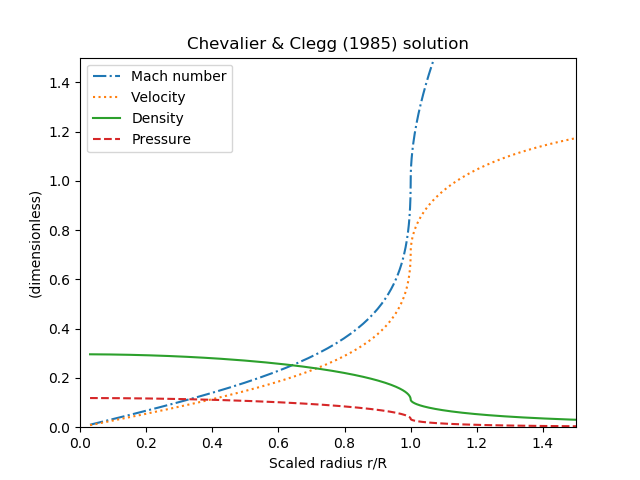}
\caption{Steady-state wind solution by \citet{CC85}. Shown are the
dimensionless quantities given in eqs.~(\ref{eq:ustar}-\ref{eq:pstar}).
Section~\ref{sec:sswinds} for details.}
\label{fig:cc85sol}       
\end{figure}

Given these conditions, the 1D hydrodynamics equations can be solved
analytically \citep[see also][]{Zhangea14}. Pressure, density
and outward velocity are given by 
\begin{equation}
    \nbr{\begin{array}{c}
         p \\ \rho \\u \end{array} } =
         \nbr{\begin{array}{c}
         p_* \dot{M}^{1/2}\dot{E}^{1/2} R^{-2}\\ 
         \rho_* \dot{M}^{3/2}\dot{E}^{-1/2} R^{-2}\\
         u_* \dot{M}^{-1/2}\dot{E}^{1/2}\end{array} }\, ,
\end{equation}
where the functions containing the radial dependencies are given 
by\footnote{$\gamma$: adiabatic index, $5/3$ for the usual monatomic ideal gas}:
\begin{eqnarray}
    u_*^2 &=& \frac{2\M^2}{\M^2+\frac{2}{\gamma-1}}\label{eq:ustar}\\
    \rho_* &=& \frac{r_*^a}{4 \pi u_*}\\
    p_* &=& \frac{2\rho_*}{\gamma\nbr{\M^2+\frac{2}{\gamma-1}}}\label{eq:pstar}
\end{eqnarray}
with $r_*=r/R$, $a=1$ (-2) for $r_*<1$ $(r_*>1)$, and the implicit 
definition of the Mach number $\M$:
\begin{equation}
    r_* = \left\{\begin{array}{lr} \nbr{\frac{\gamma-1+2\M^{-2}}{\gamma+1}}^\frac{\gamma+1}{2+10\gamma} \nbr{\frac{3\gamma+\M^{-2}}{1+3\gamma}}^{-\frac{3\gamma+1}{5\gamma+1}} & r_*<1\\ \nbr{\frac{\gamma-1+2\M^{-2}}{\gamma+1}}^\frac{\gamma+1}{4\gamma-4} \M^\frac{1}{\gamma-1}& r_*>1\end{array}\right. 
\end{equation}
The solution is shown graphically in Fig.~\ref{fig:cc85sol}.

The \citet{CC85} solution is characterised by a slow, hot and subsonic
flow inside the star cluster. The flow turns supersonic at the boundary
of the source region and then continues to accelerate towards an asymptotic
value of $1.414 \sqrt{\dot{E}/\dot{M}}$. 

As an example we give here parameters
for the Arches cluster, one of the most massive, young
\citep[2-3~Myr,][]{Lohrea18} star clusters 
in the Milky Way. 
\citet{ClarkJea18} estimate $\gtrsim50$ stars with
masses $\gtrsim 60M_\odot$. Using the initial mass function from 
\citet{Kroupea13}, this translates to a total mass of $5\times 10^4 M_\odot$
\citep[consistent with the kinematic measurement,][]{Clarksea12},
and a total number of massive stars 
$>8 M_\odot$ of $\approx 1000$ \citep[compare also][who further give a cluster radius of 0.2~pc]{Figea99,Figea02}.
Population synthesis of stellar mass loss and energy output then yields
\citep{Vossea09}: $\dot{M}\approx 10^{-3}M_\odot$~yr$^{-1}$ 
and $\dot{E}\approx 2\times10^{39}$~erg~s$^{-1}$ \citep[similar estimates can be found in][]{SH03}.
Within the cluster this yields particle densities and temperatures of
\begin{eqnarray}
n_0 &=& 300 \, \mathrm{cm}^{-3} \left(\frac{\dot{E}}{2\times 10^{39}\,\mathrm{erg}\,\mathrm{s}^{-1}}\right)^{-1/2}
\left(\frac{\dot{M}}{10^{-3}\,M_\odot\,\mathrm{yr}^{-1}}\right)^{3/2}
\left(\frac{\dot{R}}{0.2\,\mathrm{pc}}\right)^{-2}\\
T_0 &=& 9\times 10^7 \, \mathrm{K} \left(\frac{\dot{E}}{2\times 10^{39}\,\mathrm{erg}\,\mathrm{s}^{-1}}\right)
\left(\frac{\dot{M}}{10^{-3}\,M_\odot\,\mathrm{yr}^{-1}}\right)^{-1}
\end{eqnarray}
Arches and similar young, massive and compact clusters are therefore expected to be faint
diffuse X-ray emitters \citep[e.g.,][]{CRR00,ATS09}, 
which has been confirmed by X-ray observations
for the Arches cluster, Westerlund~1 and possibly 
also the Quintuplet cluster
\citep{Yusea02,WangQea06,KNM11}. Generally the temperature is somewhat lower
than predicted by the \citet{CC85} model \citep{SH03}. This may be
related to unaccounted for effects of non-equilibrium ionisation \citep{JWK06}
or limitations of our understanding of mass loading and thermalisation
efficiency of the winds. Also, the spatial distribution of the massive stars plays a role.
X-ray emission is also expected 
from the interaction of the cluster wind with the surrounding gas. The superbubble is particularly
bright in soft, $\approx 1$~keV, X-rays, 
whenever individual supernova
shock waves interact with the shell \citep{Krausea14a}.
Cluster winds can, however, be identified by their
harder spectra and co-location with the optical star
cluster \citep{Siliea05}.

The basic wind solution has been modified and enhanced by many authors
for example to include cooling \citep{Siliea04} and more sophisticated 
shapes for the source functions
\citep{Silichea11,Palea13,Zhangea14}. 
\citet{Wuea11} present detailed models for the entire
phase when massive stars ($>8M_\odot$) are present in the cluster.
They find that clusters are always in a steady outflow
regime similar to the Chevalier-Clegg model, unless the energy
input is significantly overestimated (factor $\gtrsim20$) 
by current population synthesis models or the wind loads
a significant amount of gas that was leftover from the
star formation event.
If those conditions applied, part of the massive star
ejecta would cool, be compressed by the remaining hot
gas into UV-shielding filaments and form stars in an extended
or second star formation episode \citep{PWT14}.
The total cold gas dropout from the wind can reach 1-6\% of 
the total stellar mass of a cluster \citep{WPT17}.
The population of stars formed would be 
(moderately due to the mass loading) enriched in He-burning
products ejected from Wolf-Rayet stars and supernovae
\citep{Wuea11}.
The latter two predictions disfavour this mechanism 
as explanation for the frequently observed chemically
distinct populations in globular clusters (compare \S\ref{sec:nucsy}).

For a normal stellar population, type Ia supernovae
appear from about 100~Myr after a star formation event
at a rate characterised by the delay time distribution
\citep{Heringea19}:
\begin{equation}
    \mathrm{DTD}(t) = 7\times 10^{-13} M_\odot^{-1}\,\mathrm{yr}^{-1} \nbr{\frac{t}{\mathrm{Gyr}}}^{-1.34}\, .
\end{equation}
For a $10^6 M_\odot$ cluster at an age of 100~Myr, 
this yields about 15 events
per Myr. This is near the limit where one might consider
the energy injection continuous and apply cluster wind models.
\citet{DErcolea08} show in a 1D hydrodynamic simulation
with individual SN~Ia that even one event can turn the cluster
into an outflow state. A SN~Ia rate comparable to the
one in the field would leave star clusters in a 
continuous outflow state.

SN~Ia occur in binary systems
\citep[e.g.,][]{Diehlea14a}, which may be more frequent 
in massive star clusters \citep{Leighea15}.
Direct searches for type Ia SNe in massive star clusters
have, however, so far only produced upper limits
\citep{WB13}. Dynamical effects should lead to a high 
net destruction rate for binaries in clusters, so that,
at least at late times,
they may have actually fewer SN~Ia than the field
\citep{Chengea18,Bellonea19}.

\subsection{Cooling flows?}\label{sec:coolflows}

After the end of the type~II supernova phase 
($\approx 30-40$~Myr after star formation) and before SN~Ia
start to occur \citep[$\approx 100$~Myr, e.g.,][]{LiuStan18} 
a star cluster has little internal energy
production and may in principle have a cooling flow.
\citet{DErcolea08} show that mass loss and energy
injection are dominated by AGB stars:
\begin{eqnarray}
\alpha &=&\dot{M}/{M}=3\times10^{-17}\,\mathrm{s}^{-1} \\
\dot{E}/{M} &=& \alpha v_\mathrm{w}^2/{2}=
10^{29} \,\mathrm{erg}\,\mathrm{s}^{-1}\,\mathrm{M_\odot}^{-1}
\end{eqnarray}
The cooling flow is robust for these parameters as they also
show that more than ten times higher energy input would 
be required to turn the cooling flow into a wind.

\citet{DErcolea10,DErcolea12,DErcolea16} and others have
argued that the gas mass flowing to the centre 
may initiate secondary star formation. Challenges 
in explaining the chemically distinct multiple populations
in globular clusters (compare \S\ref{sec:nucsy}) in this
way include the small mass of enriched material available
and fine-tuning required for the dilution of the ejecta.

\citet{CS11} have conjectured that the gas in the
cooling flow would actually not be able to form stars for 
several 100~Myr, because the UV flux of the remaining
intermediate-mass stars would keep the gas photo-dissociated 
and too warm for star formation. The accumulating gas would 
only form stars when the UV luminosity has declined enough 
to allow the formation of molecular hydrogen.
However, gas cooling can also take place very efficiently
in atomic gas via $C^+$ \citep{GlovClark12}. Also, it is
unclear if type~Ia SNe would be delayed sufficiently for the
model to work \citep{Lymea18}.

Dense gas or late star formation as 
postulated in the above cooling 
flow models is generally not observed in star clusters
\citep[e.g.,][]{CabZiea15,Longmore2015,BL18}.
This calls into question our understanding of the gas 
dynamics in star clusters with ages between the
type~II and type~Ia supernova phases. One possibility
is that the cooling flow gas accretes on to 
the dark remnants, i.e., the stellar mass black holes and 
neutron stars \citep{Krausea13b,RK19}. The energy
released in jets, winds and radiation could then 
drive a cluster wind. \citet{DErcolea08} derive a critical
energy input of $6 \times 10^{37}$~erg~s$^{-1}$
for their $10^7 M_\odot$ cluster.
Even the typical
luminosity of one X-ray binary 
\citep[few $10^{38}$~erg~s$^{-1}$,][]{Jordea04}
would be sufficient to accomplish this. 
Pulsar winds can keep star clusters in an outflow state \citep{Naimea20}.

\section{Collisional dynamics and long-term evolution}\label{sec:nbody}

After the gas cloud 
a star cluster formed from has been partially transformed into stars and  dispersed, its fate  
 is governed
by gravity (i.e. collisional dynamics and  tidal perturbations) and mass loss of the stars due to stellar evolution. Here we discuss the various physical processes separately, but it is important to keep in mind that most processes act simultaneously and an important area of research is understanding the interplay between them, which is often non-linear. 

\subsection{Stellar evolution}
Stellar evolution leads to a  decrease of the total cluster mass, at a rate that is slow compared to the orbital frequencies of the stars, such that the cluster can approximately maintain its virial equilibrium. The removal of mass leads to a reduction of the binding energy and an increase of the cluster radius. 
 If the stellar mass loss happens throughout the cluster with no preferred location then the cluster radius is inversely proportional to the mass
 \citep{Hills1980}. For mass segregated clusters most stellar mass loss occurs in the cluster centre where the binding energy is larger, resulting in a faster expansion.

\subsection{Tidal shocks}
{During the first 0.1-1~Gyr, cluster dissolution is likely dominated by tidal `shocks', i.e. impulsive tidal perturbations from Galactic substructure, such as transient spiral arms and molecular gas clouds \citep[e.g.][]{2006MNRAS.371..793G,2010ApJ...712..604E,Kruijea11}.  These perturbations boost the energy of stars in the cluster, some of which will exceed the escape energy and will therefore become unbound \citep{1958ApJ...127...17S}. 
The rate of shock-driven mass loss scales inversely with the mass volume density of the cluster, and is proportional to the surface density of the individual clouds and the ISM density in the host galaxy disc \citep{1958ApJ...127...17S}.

When integrated over the lifetime of a cluster, this mass loss mechanism could dominate the total mass loss budget \citep{2010ApJ...712L.184E,2015MNRAS.454.1658K}, even in environments of relatively low gas density such as the solar neighbourhood \citep[e.g.][]{1958ApJ...127...17S,2006MNRAS.371..793G,2006A&A...455L..17L}, where a single encounter with a GMC ($\gtrsim10^5~\msun$) can completely disrupt a modest open cluster \citep[$\sim10^3~\msun$,][]{Wielen1985,1987MNRAS.224..193T}.  
By scaling $N$-body models of individual encounters, it was found that in gas-rich environments like galaxy discs, GMCs dominate the disruption of clusters \citep{2006MNRAS.371..793G,Webbea19},
decimating the initial globular cluster population to the survivors that remain at the present day \citep{2010ApJ...712L.184E,2015MNRAS.454.1658K}.

Tidal shocks do not only drive considerable mass loss, they also dominate the  structural evolution of stellar clusters: after an initial phase of expansion due to the escape of unbound stars \citep{Webbea19},  the remaining cluster of bound stars may shrink due to energy conservation \citep[centrally concentrated clusters shrink, while low-concentration clusters expand, see][]{2016MNRAS.463L.103G}. When ignoring other effects, a density increase makes tidal shocks self-limiting \citep{1999ApJ...513..626G}. However, a higher density makes  two-body relaxation more important which tends to reduce the cluster density, thereby counteracting the shock-induced density increase. Under the assumption that statistical equilibrium is reached, eventually the ratio of the shock dissolution time-scale and  the relaxation time-scale will become constant, resulting in a shallow mass-radius relation \citep[$\rh\propto M^{1/9}$,][]{2016MNRAS.463L.103G}. The normalisation of the predicted mass-radius relation depends on the environment, such that clusters are smaller at higher ISM densities (which is likely already the case at formation, see \citealt{choksi19}), slowing down their shock-driven disruption. However, even for correspondingly more compact clusters, \citet{2015MNRAS.454.1658K} predict that the total shock-driven mass loss dominates over relaxation-driven mass loss when considering the dynamical evolution of globular clusters over a Hubble time.

To obtain a complete understanding of the interplay between shocks and relaxation, a comprehensive parameter study of $N$-body simulations including both processes is required, which again highlights that this is an important area for future research. A complementary approach to  controlled $N$-body experiments would be to use direct $N$-body simulations with realistic particles numbers ($N\gtrsim10^6$) and evolve them in the time-dependent tidal field extracted from models of galaxy formation  at the epoch of GC formation.

\subsection{Two-body relaxation}
\label{sec:2bdrel}
The importance of collisional dynamics in the evolution of star clusters depends on the evolutionary stage of the cluster and the timescale that is considered. Given sufficient time, all clusters will dissolve due to collisional effects, even the clusters that are not in a Galactic tidal field \citep{2002MNRAS.336.1069B}.

Globular clusters are the archetypical collisional systems, that survived the initial phase of tidal shock-driven disruption, meaning that orbital energy diffusion via gravitational interactions -- so-called two-body relaxation, or collisional dynamics -- plays an important role in their evolution. This is because the velocities of stars are relatively low ($\sim10$~km/s) and stellar densities are high ($\sim10^{4-6}~{\rm pc}^{-3}$), making two-body encounters frequent and long-lasting. Another way of saying this is that the relaxation timescale is short  (few Gyr) compared to their ages (10-12~Gyr). 
{Two-body relaxation is also relevant during the formation phase of clusters, contrary to some propositions made in the literature \citep[e.g.][]{2001ApJ...561..751F, 2014prpl.conf..243K}.}
To explain this, we start by painting a broad-brush picture of the classical theory of relaxation that was developed for (old) globular clusters.

The consequences of two-body relaxation are reasonably well understood for the idealised case of a single-mass cluster, which is often regarded to be a reasonable approximation for globular clusters, because their stars are confined to a narrow   range of masses. A single-mass cluster,  
with stars
initially  in hydrostatic equilibrium, without primordial binaries, develops a radial energy flow  as a result of  energy diffusion such that energy flows through the half-mass radius (\rh) at a rate $\sim |E|/\trh$ (ignoring constants of order unity).  Here $E\sim -GM^2/\rh$ is the total energy of the cluster, with $G$ the gravitational constant and $M$ the total cluster mass. The timescale $\trh$ is the half-mass relaxation time, which we shall define below (equation~\ref{eq:trh}). 
This energy flow originates from the core, where stars lose kinetic energy to the stars outside the core via two body interactions. As a result, the velocity dispersion of the stars in the core reduces -- i.e. they `cool' -- and the core radius contracts, while the stars outside the core heat up. The stars in the core now experience a higher binding energy and due to the virial theorem,  the stars now move faster. This somewhat paradoxical result is a direct consequence of the negative heat capacity of self-gravitating systems. The time evolution of the cluster structure can be solved in various ways. \citet{1980MNRAS.191..483L} used a set of equations that are strikingly similar to the stellar structure  equations (so-called gaseous models, or continuum models) and found that this process of core contraction continues until the core has an infinite density and zero mass: the core has collapsed. In reality this mathematical endpoint is never reached, and a binary star forms when the number of stars in the core has reduced to $\sim10$ \citep{1987ApJ...313..576G}. To sustain the two-body relaxation process after core collapse, the freshly formed binary absorbs the negative energy that continues to flow into the core, by hardening in interactions with other stars.  From that moment onwards, the evolution of the cluster is approximately self-similar, with the rate of energy absorption  of the central binaries determined by the global energy flow through the cluster \citep{H61}

\begin{equation}
\Edotbin \propto -\frac{|\Eext|}{\trh}.
\end{equation}
Here $\Eext$ is the external energy of the cluster, which is the total energy excluding the negative energy locked up in binaries (i.e. $\Eext \sim -GM^2/\rh$). Because of energy conservation, the external energy must increase at a rate 
\begin{equation}
\Edotext = -\Edotbin.
\label{eq:Edotext}
\end{equation}
The insight that there must exists a balance between the rate of energy production in the core and the energy flow through the cluster came from Michel H\'{e}non \citep{1975IAUS...69..133H} and is a fundamental building block in the theory of cluster evolution. It allowed him to derive two models for the post-collapse evolution: in the absence of a Galactic tidal field, the energy increase leads to an expansion of the cluster at an approximately constant mass \citep{H65}, while tidally limited clusters lose mass over the tidal boundary (sometimes referred to as `evaporation') at a constant rate, while maintaining a constant density \citep{H61}. These two solutions describe the two extreme ends of the life cycle of  tidally limited star clusters with high initial density. By smoothly `stitching' the two models the relaxation driven evolution of star clusters can be described from {the moment they emerge from a gas-rich environment (i.e.\ once tidal shocks no longer dominate the instantaneous disruption rate)} to their eventual dissolution \citep{2011MNRAS.413.2509G}. Thanks to its analytic  nature,  this simple model for the relaxation driven evolution of star clusters readily provides expressions for $M(t)$ and $\rhoh(t)$ at different Galactocentric radii, which can be used to construct evolutionary `tracks' and `isochrones' of globular cluster radius (or density) and mass as a function of location in the Galaxy,
which -- despite their first order nature -- provide a satisfactory match with the {observed mass-density distribution of globular clusters} (see Fig.~\ref{fig:rhom}), supporting H\'{e}non's suggestion that collisional dynamics is important for almost all globular clusters and {shapes these relations.}

\begin{figure}\centering
  \includegraphics[width=0.55\textwidth]{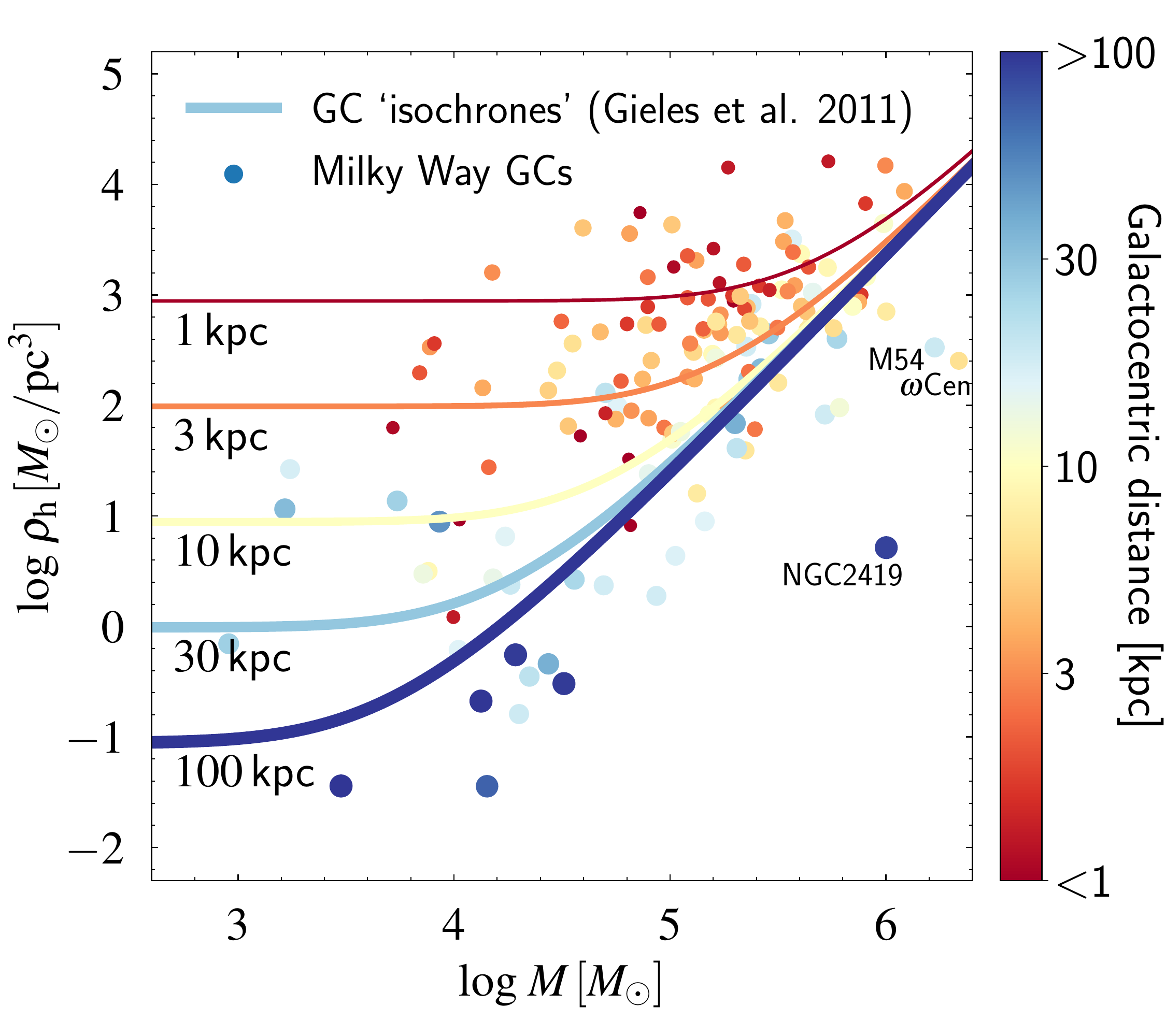}
\caption{The half-mass density of Milky Way globular clusters as a function of their mass ($x$-axis) and Galactocentric radius (colour coding). `Isochrones' from the cluster evolution model of \citet{2011MNRAS.413.2509G} are shown. 
This diagram is the equivalent of the Hertzsprung-Russell diagram of stars. Collisional dynamics is not yet important for objects that are well below the model lines.}
\label{fig:rhom}       
\end{figure}

\subsection{Two-body relaxation in young clusters}\label{sec:2bd-rex}

We now turn to the relevance of relaxation for young clusters. To get an idea for the physical time it takes for relaxation to become important, we write the expression for $\trh$ from \citet{1971ApJ...164..399S} as
\begin{equation}
\trh \simeq 18~\myr~\psi^{-1} \frac{M}{10^4~\msun}\left(\frac{\rhoh}{10^4~\msun/{\rm pc}^3} \right)^{-1/2},
\label{eq:trh}
\end{equation}
where $\rhoh=3M/(8\pi\rh^3)$ is the average mass density within $\rh$. 
To derive this, we assumed an average mass of stars of $0.5~\msun$ and assumed that the slight dependence of the Coulomb logarithm on the number of stars can be neglected ($\ln\Lambda = 8$). The term $\psi\ge1$ is a factor that depends on the mass spectrum within $\rh$, and often the assumption of single-mass clusters is made (i.e. $\psi=1$). 
It takes about 16 initial $\trh$ for a single-mass cluster to reach core collapse \citep[e.g.][]{1980ApJ...242..765C}, which even for the relatively low-mass and high-density scaling adopted in equation~(\ref{eq:trh}) corresponds to a relatively long time of $\sim300~\myr$, i.e. well after star formation ceased, suggesting that relaxation plays no role in the early evolution of clusters.

However, there are several important differences between young clusters and their older counterparts that are important for this discussion and make  collisional dynamics important at young ages.  Most importantly, young  clusters have a well populated (initial) stellar mass function between $\sim0.1~\msun$ and $\sim100~\msun$. The presence of the high-mass stars significantly speeds up the relaxation process,  because the energy transfer from high-mass stars to low-mass stars is more efficient than between stars of the same mass\footnote{The single-mass approximation, therefore, also breaks down for globular clusters with a stellar-mass black hole population \citep{2013MNRAS.432.2779B, 2019MNRAS.487.2412G, 2019arXiv191100018K,Wang2019,2019arXiv190611855A}.}. With numerical $N$-body experiments, \citet{2002ApJ...576..899P} find that core collapse happens after $0.2\trh$ (with $\trh$ defined with $\psi=1$)
, i.e. 2 orders of magnitude faster than for single-mass clusters \citep{2010MNRAS.408L..16G}, which corresponds to $\sim3~\myr$  for the fiducial case in equation~(\ref{eq:trh}), and earlier for lower-mass/denser clusters. 
The definition of core collapse is often taken to be the moment of dynamical formation of the first hard binary, which means that in the first 3~Myr the most massive stars are migrating to the centre of the cluster (if they did not  form there), making collisional dynamics important already before core collapse. 

There are several reasons to assume that binary activity starts even earlier. 
Firstly, clusters may 
form hierarchically from mergers of lower-mass and denser sub-clusters, which can each dynamically form a  binary. To quantify the importance of this, lets assume that the cluster formed from two clumps with each half the mass of the cluster. If they merge with negligible orbital energy, we find from conservation of energy that the clumps have half the radius of the final cluster, i.e. 4 times shorter relaxation time, such that the clumps undergo core collapse within a Myr. Considering an additional step in the hierarchy would reduce $\trh$ of the first sub-clumps to form to $\sim10^5~$yr, i.e. well within the timescale of cluster formation itself. In addition, the massive stars may form in the centre of the sub-clumps/cluster (see Section~\ref{sec:mseg}), resulting in a cluster forming in a core collapsed state,  setting off binary activity immediately.  Finally, several models of the formation of star clusters  \citep{Longmea14, VS+17, 2018MNRAS.478.2461G, 2019arXiv190901565K} and massive stars  \citep{Padoan+2019} suggest that gas inflow from larger scale is important (compare \S\ref{sec:frag}). Accretion of low-angular momentum gas on proto-stars leads to an efficient contraction of the parent cluster \citep{1998MNRAS.298...93B}, driving the cluster into core collapse by reducing $\trh$ as the cluster gains mass \citep{2011MNRAS.410.2799M}.  

\subsection{Mass segregation in young clusters} \label{sec:mseg}

Observational support for the importance of collisional dynamics at young ages comes from the (low) densities of young-stellar objects in the solar neighbourhood, which is consistent with the density distribution of a population of dynamically expanding (i.e. post-collapse) low-mass star clusters \citep[few 100 stars,][]{2012MNRAS.426L..11G}.

Collisional dynamics will equalise ``temperature" between stars and hence the kinetic energy per star. More massive stars will therefore acquire lower velocities to have the same kinetic energy as the lower mass stars. The high-mass stars will therefore be found deeper in the potential well, i.e.,
closer to the centre of the cluster. This is called mass segregation. Dynamical mass segregation is a consequence of two-body relaxation, whereas primordial mass segregation describes a situation where the massive stars already form in the centre.

The observational assessment of mass segregation is mixed.
Young star clusters show signs of mass segregation \citep{Hillenbrand98, deGrijs02, LittlefairEtAl2003, StolteEtAl2005, Stolte06, KimEtAl2006, HarayamaEtAl2008, EspinozaEtAl2009, BontempsEtAl2010, GennaroEtAl2011}. However, observations of pre-main sequence stars in star-forming regions do not indicate mass segregation \citep{ParkerEtAl2011, ParkerMaschbergerAlves2012, GennaroEtAl2017, ParkerAlves2017, DibSchmejaParker2018}. \citet{PlunkettEtAl2018} find the prestellar cores in Serpens South to be 
mass-segregated, whereas the pre-main sequence stars are not. The complexity of defining the details of clusters and subclusters in combination with measurement differences complicate the problem of the distribution of masses. \citet{KirkMyers2011} observe mass segregation in small groups in Taurus. Investigating the stellar distribution in total, \citet{ParkerEtAl2011} find the most massive stars to be inversely mass segregated. 

\citet{ParkerDaleErcolano2015} point out that the degree of mass segregation depends on how it is measured in simulations and compared to observations.

It
can conveniently be computed using the method of \citet{Allison09, Allison09a}, which is based on the minimum spanning tree \citep{GowerRoss69}. 
\citet{Kirk+2014} showed that small clusters in hydrodynamic simulations exhibit primordial mass segregation with distributions consistent with nearby, young embedded clusters.  
\citet{ParkerDaleErcolano2015} found that the degree of mass segregation is reduced if the clusters form under the influence of feedback  
from massive stars. 
\citet{GirichidisEtAl2012b} report that the degree of mass segregation depends on the initial density configuration, but that no inverse mass segregation occurred. 
Clusters undergoing competitive accretion are expected to be primordially mass-segregated \citep[cf.\citeauthor{Bate2009} \citeyear{Bate2009}]{Bonnell01}.
In all cases the time scales are consistent with dynamical relaxation times, so all clusters had enough time to dynamically mass segregate. 

\subsection{Dynamical feedback}
We conclude this section by discussing the feedback from collisional dynamics on star formation. 
The dynamically formed binaries consist of massive stars, which soon after formation start ejecting other (massive) stars in dynamical interactions \citep[see also \citeauthor{Gavagnea17} \citeyear{Gavagnea17}]{1967BOTT....4...86P},
possibly explaining the origin of  the O-stars that are found with high velocities ($\gtrsim30~$\kms), far from star forming regions \citep{1961BAN....15..265B}. In addition, a large fraction if not all massive stars are expected to form in binaries and higher order multiples \citep[e.g.,][]{2012Sci...337..444S}, which have larger gravitational cross section than single stars making binary-binary interactions an additional channel for ejecting massive stars from 
an ongoing cluster formation site \citep{1990AJ.....99..608L}. The removal of massive stars reduces the mechanical and radiation feedback from massive stars on the cluster and the more distributed feedback in the low(er) density ISM has consequences for galaxy formation \citep{2009ApJ...695..292C}. Finally, the high central density of massive stars in the centre affect the 
ionisation level (and thus accretion rate)
and survival of discs around smaller mass stars

In conclusion, collisional dynamics is likely important from the very beginning of cluster evolution and it may have played a role in the origin of the multiple populations in GCs (see \S\ref{sec:SMS}), with tidal perturbations being an additional important process in the early evolution. It is unclear, if observed signs of mass segregation are of dynamical or primordial origin. Mass segregation and ejection of massive stars modify ionisation levels in accretion discs and hence accretion rates on to stars and the strength of feedback from star clusters.

\section{Nucleosynthesis}\label{sec:nucsy}

For a long time, nucleosynthesis (or, in other words, internal chemical evolution) has been ignored in star cluster modelling, based on both theoretical and observational arguments. Galaxies have a deep potential well and are hence expected to retain even some of the ejecta that massive stars shed at high velocity. This has recently been confirmed by measurements of Doppler kinematics of the radioactive decay line of unstable $^{26}$Al, which traces high-mass star ejecta \citep{Kretschea13,Krausea15a,2019MNRAS.490.1894R}. The high observed velocities suggest that a large fraction of the ejecta is blowing away from their birth places at high speeds. The scale height of the order of kpc is in agreement with expectations from fountain-flow super-bubbling disc models where ejecta diffuse into the hot halo and return in part on a Gyr timescale \citep{2019A&A...632A..73P,2019MNRAS.490.1894R}. 

The need for a sufficiently deep potential well to retain the gas despite the energetic feedback from the massive
stars and eventually recycle it internally to make new stars is supported by the fact that open clusters present no (within measurement uncertainties) spread in Fe-peak, $\alpha$, and s-process elements (hereafter heavy metals). These specific species actually vary only in the most massive globular clusters (hereafter GCs), with the extreme case being Omega Cen which is thought to be the remnant of a dwarf galaxy nucleus \citep{Butler+78,Zinnecker+88,2011ApJ...731...64M}. 
Such rare objects (recently called Type II GCs; see e.g. \citealt{2017MNRAS.464.3636M} and \citealt{2018ApJ...859...81M}) possibly make the link between star clusters (open clusters and Type I GCs) and chemically evolved dwarf galaxies. 

Interestingly, large variations in carbon, nitrogen, oxygen, sodium, magnesium, and aluminium (C, N, O, Na, Mg, Al, hereafter light elements) were discovered in GCs already in the 1970's (among bright red giants, \citealt{1971Obs....91..223O}; see \citealt[][and references therein]{Kraft1979}), but they were initially attributed to internal deep mixing processes occurring along the evolution of the stars themselves (\citealt[e.g.][]{1979ApJ...229..624S,1990SvAL...16..275D,1993PASP..105..301L,1996ApJ...464L..79C,1996A&A...308..773D,1997IAUS..189..193D,2000A&A...356..181W}; but see e.g. \citealt{1980ApJ...237L..87P} and \citealt{1992AJ....104.1818B} who already advocated for a primordial origin). Hence, the ``classical paradigm" became established, presenting individual GCs as the archetype of a single, coeval, and chemically homogeneous stellar population, i.e., a system that did not undergo any internal chemical evolution.

\subsection{Light element abundance variations}
The surprise came in the 2000's from studies with 8-10m class telescopes which opened a spectroscopic window on less evolved stars down to the main sequence turnoff in the case of the closest GCs \citep{2001A&A...369...87G,2001A&A...373..905T}. All the Galactic and extra-galactic GCs that were scrutinised this way were shown to host multiple stellar populations (hereafter MSP) located all along the color-magnitude diagram and exhibiting similar variations in light elements \citep[for reviews see][]{GCB12,Grattea19,Charb16,BL18}. 
Extensive surveys established that while every Galactic and extra-galactic GC contains main sequence and red giant stars with field-like abundances (the so-called first population, 1P), 30 to 90 \% of their hosts (the second population, 2P) actually line up along an O-Na anticorrelation \citep[e.g.][]{2009A&A...505..117C,2009A&A...505..139C,Carea10a,2009A&A...503..545L,GCB12,2016A&A...592A..66W}, a C-N anticorrelation \citep[e.g.][]{1979ApJ...230L.179N,1985ApJ...299..295N}, and a Mg-Al anticorrelation with a possible linkage towards Si depletion in the case of the most massive and/or most metal-poor GCs \citep[e.g.][]{1981ApJ...244..205N,1999AJ....118.1273I,2003A&A...402..985Y,2015ApJ...810..148C,2015AJ....149..153M,2019MNRAS.tmp.3134M,2017A&A...601A.112P,2018ApJ...859...75M,2019A&A...622A.191M}. 
This definitively 
argued
for a primordial origin of these anomalies, 
calling for GC self-enrichment by the hot hydrogen-burning yields of short-lived massive stars before or during the formation of the low-mass stars that we observe today \citep{2007A&A...470..179P,2017A&A...608A..28P}.
The  light  element  abundance
variations within GCs reflect in the photometric properties of their MSPs. 
When one uses  specific  combinations  of  optical  and  ultraviolet  HST filters,
the MSPs 
spread out 
along broadened or multiple sequences of the color-magnitude diagram (CMD) and of the so-called chromosome map 
\citep[e.g.][compare Adamo et al. 2020, in prep.]{2004ApJ...605L.125B,2007ApJ...661L..53P,2012ApJ...760...39P,2015AJ....149...91P,2008A&A...490..625M,2019MNRAS.487.3815M,2009ApJ...707L.190H,2015MNRAS.447..927M,2017MNRAS.464.3636M,2017ApJ...844..164B,2019MNRAS.487.3239Z}. 

\subsection{Multiple sequences in the colour-magnitude diagram}
Color variations and/or separations in the CMD are used to infer He abundance variations among MSPs (typically between 0.003 and 0.19 in mass fraction, with He enrichment increasing with the present-day mass; \citealt[e.g.][]{2004ApJ...612L..25N,2005ApJ...621..777P,2012AJ....144....5K,2011A&A...534A...9S,2015MNRAS.446.1672M,2015MNRAS.451..312N,2018MNRAS.481.5098M,2019ApJ...871..140L}). Importantly, the photometric approach revealed the presence of
MSPs similar to GC ones in extragalactic massive star clusters with ages down to $\sim$2~Gyr \citep{2014ApJ...797...15L,2016EAS....80....5B,2016ApJ...829...77D,2017MNRAS.464...94N,2017MNRAS.465.4159N,2018MNRAS.477.4696M,2019MNRAS.487.5324M,2019MNRAS.486.5581G,2019MNRAS.485.3076N}, with the extent of the MSP increasing with the age of the clusters \citep{2019svmc.confE..11M}. It thus seems that the formation of MSPs was not restricted to old GCs, but that it continued to occur in sufficiently massive star clusters 
down to a redshift of $\sim$0.17. This provides a very strong link between YMSC and ancient GCs, and suggests a common formation and evolution path.

The exact shape and the extension of these features vary from GC to GC, and depend primarily on their mass, compactness, metallicity, and age \citep[e.g.][]{2009A&A...505..139C,Krausea16a,2017A&A...601A.112P,2019A&A...622A.191M,2019svmc.confE..14C,2019svmc.confE..11M}, and these abundance patterns have never been found in open clusters \citep{2010A&A...511A..56P,2012A&A...548A.122B} with the possible exception of NGC~6791 (\citealt{2012ApJ...756L..40G}, but see \citealt{2014ApJ...796...68B}).

\subsection{Self-enrichment scenarios}
Different self-enrichment scenarios were proposed to explain the abundance patterns described above.

\subsubsection{AGB model}
In the `AGB model' 
\citep[e.g.][]{2001ApJ...550L..65V,DErcolea12,2013MNRAS.431.3642V,2016MNRAS.458.2122D}
it is assumed that a second generation of stars forms from material that is polluted by AGB winds from a first generation. 
The model starts from the point that AGB winds are slow enough so that they may be unable to escape the potential well of a massive star cluster. It is conjectured that
after the type~II supernova phase, AGB winds would be 
the only energy source for the intracluster gas, which would be insufficient to overcome radiative losses.
Consequently, a cooling flow forms, and stars would form 
in the centre of the cluster.
This scenario has been criticised in multiple ways:
AGB nucleosynthesis builds an O--Na
correlation instead of the observed anti-correlation
\citep{1997A&AS..123..241F},
and it releases He-burning products, thus predicting total C+N+O
variations that are only observed in a few GCs \citep[e.g.][]{2009A&A...505..727D,2015MNRAS.446.3319Y}.
The AGB stars need to be massive
enough to undergo hot-bottom burning to reach the required temperatures
\citep[$\sim6.5~\msun$,][]{2001ApJ...550L..65V},
which implies that not enough mass is available to pollute (in some cases) $\gtrsim80\%$ of the stars. 
This is commonly referred to as the mass budget problem 
\citep[e.g.][]{2006A&A...458..135P,2011MNRAS.413.2297S,GC19}.
Some of the ideas that have been put forward to overcome the mass budget problem, such as the loss of $\gtrsim90\%$ of the first generation stars over the tidal boundary \citep{DErcolea08}, do not address the GC mass dependence of the fraction of polluted stars, i.e. the requirement of more polluted material per unit of  GC mass in more massive GCs. We refer to this as the specific mass budget problem. 
Whether star clusters have a cooling flow phase at all is questionable (\S\ref{sec:coolflows}). Also, the gas expulsion paradigm, which had been identified as the mechanism to expel the surplus 1P stars, may not 
be applicable to many low-mass clusters (\S\ref{sec:gex},\S\ref{sec:obskinYSCs}), and would certainly require some very unusual assumptions
for it to work in high-mass clusters (\S\ref{sec:gasexpul}, \citealt{2015MNRAS.453..357B}).
Finally, in dwarf galaxies the amount of field stars with the same metallicity is comparable to the total mass in GCs, putting an upper limit of $\lesssim50\%$ on the amount of mass that GCs could have lost  \citep{2012A&A...544L..14L}. This is referred to as the external mass budget problem. Recent models combining the variety of dynamical mass loss mechanisms discussed in Section~\ref{sec:nbody} predict that globular clusters were only 2--4 times more massive at birth \citep{2015MNRAS.454.1658K,Reinaea18}. 

\subsubsection{Fast-rotating massive star model}
Ordinary massive stars ($\sim20-100~\msun$) have the correct central temperature to create the O-Na anti-correlation, but they are not convective and the material does therefore not reach the surface. To overcome this, \citet{2007A&A...464.1029D} assume fast-rotating massive stars (FRMS) which undergo rotationally induced mixing 
and possibly lose a large amount of material through a mechanical wind via an outflowing (decretion) disk.
\citet{Krausea13b} laid out a detailed scenario,
showing that 2P stars may form in the decretion discs
from material spreading through the disc from the
surface of the star, and accreting pristine gas 
during a somewhat extended embedded phase of $\approx 10$~Myr. The embedded phase would be longer in massive
star clusters, because stellar feedback would not be 
strong enough to remove the gas.
However, the FRMS scenario 
also faces the mass budget problem \citep{2006A&A...458..135P,Krausea13b}.
In addition, FRMS reach high-enough temperatures to activate the MgAl chain only when the He fraction has increased significantly, predicting a larger He spread in GCs with a Mg--Al anti-correlation than observed 
\citep[e.g.][]{2016A&A...592A.111C,2015MNRAS.451..312N,2015ApJ...808...51M}.

\subsubsection{Supermassive star model}\label{sec:SMS}
Finally, supermassive stars (SMS) have
also been suggested as polluters.
SMS models 
with masses between $\sim2\times10^3\,\msun$ and $\sim2\times10^4\,\msun$ reach the required central temperature to activate the MgAl chain ($\sim72-78$\,MK) already at the very beginning of the evolution on the main sequence, when the He abundance is close to pristine \citep{2014MNRAS.437L..21D,2017A&A...608A..28P}. Consequently, in this early evolutionary phase the H-burning products of SMSs show remarkable agreement with the various observed abundance anti-correlations and Mg isotopic ratios
\citep{2014MNRAS.437L..21D}.  A formation scenario has been proposed by \citet{2018MNRAS.478.2461G}. Similar to the picture described in detail in \S \ref{sec:frag}, 
a proto-GC would form at the confluence of gas filaments 
in gas-rich environments. At the highest density peak, 
inflow motions are converted to turbulence, and, where 
gravity dominates locally, the gas fragments into proto-stars.
Inflow over at least these two hierarchies leads to
cancellation of angular momentum by the time the gas 
reaches the proto-stars. Accretion of this low-angular momentum gas then reduces the specific angular momentum 
(i.e. the angular momentum per unit mass) and
the stellar density increases as $\rho \propto M^{10}$
\citep{1998MNRAS.298...93B}.
An SMS would then form by runaway collisions.
The SMS is assumed to be fully convective and the nucleosynthesis yields are efficiently brought to the 
surface, streaming off via the usual radiatively driven wind. The wind is initially fast $\gtrsim 1000$~\kms, but brakes down quickly by interaction with dense gas in the still embedded cluster. The ejecta would then mix into the star-forming dense gas that is accreting on the proto-stars 
in the cluster or collapse locally to form stars independently. 

{\it{If}} SMSs form via stellar collisions, then it may be possible to keep the He abundance low and also produce an order of magnitude more polluted material thereby addressing the mass budget problem \citep{2018MNRAS.478.2461G}. This model also provides a pathway to solve the specific mass budget problem because the dynamical models predict a super-linear scaling between the amount of polluted material released via the SMS wind and GC mass. As of today, SMS thus appear to be the
most appealing candidate, provided that these stars 
really exist in nature and are 
fully convective \citep[cf.][]{Haemmerlea18} so they can release the
material at the very beginning of their evolution to avoid
  overproduction of He.

It may come as a surprise that SMS would be difficult to find observationally. 
\cite{Martins_etal2020} predicted the 
detectability of cool SMS in proto-GCs at high redshift through deep imaging with JWST NIRCAM camera.
One problem at low redshift however is that clusters that would be massive enough are not found in the Milky Way. 
R136 in the Large Magellanic cloud is, at 50~kpc distance, the closest example of a young massive cluster that may just be massive and compact enough to show some signs of massive star formation via collisions. Indeed some very massive stars ($>160\msun$) have been observed
\citep{2010MNRAS.408..731C}.
Since SMS are expected to occur in embedded star clusters, absorption would likely be an issue.

An interesting alternative might be MASER emission. GHz MASERs are reliable tracers of massive star formation \citep{Ellingea18,Billea19}. SMS might hence be expected to show particularly bright MASER emission. Active Galactic Nuclei are an interesting analogue, as they also come with molecular tori of significant optical depth and strong central UV source. Strong MASER emission is frequently seen in heavily-absorbed AGN \citep{Castangea19}.
A good example is the archetypical nearby AGN in NGC~1068,
where the outer accretion disc is spatially and kinematically resolved, and the mass of the central object has thus been measured
\citep{MT97,Galea01}.

Similar to the case of AGN, forming super star clusters have also been found to be associated with 22 GHz H$_2$O MASER emission \citep{Gorskea19}. In particular, \citet{Gorskea19}
find a nuclear kilomaser in NGC~253, also associated with a forming super star cluster. The spectrum has more than one component and a total width of $\approx 170$~\kms. 
Whether this relates to SMSs in their centres, and if rotation curves of any SMS disc can be obtained, remains to be seen.

 Many other models have been proposed in the literature
 \citep{deMinkea09,Bastea13a,2017ApJ...836...80E,2018ApJ...869...35K,2019ApJ...871...20S}. A recent example is the model by Zinnecker at al. (2020, submitted).
  The model suggests that first, only convective, still accreting high-mass stars form \citep{HosoOmu09}, with slow, heavily cooling winds \citep{Vink18} producing a chemically anomalous population of predominantly low-mass stars.
  In this model, an SMS is not needed, yet it also solves the mass 
  budget problem and the He overproduction problem. 
  More research into such models is required.

\section{Synthesis: physical processes in complex systems of stars and gas}\label{sec:syn}

It is well established that a complex cycle of matter and energy takes place within galaxies.  The non-linear flow patterns in the multi-scale multi-phase interstellar medium are the central engine of galaxy evolution, they determine where and at what rate stellar clusters and associations form in our Milky Way. They build up in regions of the interstellar medium that become unstable under their own weight. These are the star forming clumps and cores in the deep interior of molecular clouds. 

Altogether, star cluster formation can be seen as a three-phase process. First, supersonic turbulence creates a highly transient and inhomogeneous molecular cloud structure that is characterised by large density contrasts. Some of the high-density fluctuations are transient. Others exceed the critical mass for gravitational contraction, they begin to collapse and eventually decouple from the complex cloud environment. Second, the collapse of these unstable cores leads to the formation of individual stars in clusters and associations. In this phase, a nascent protostar grows in mass via accretion from the infalling envelope until the available gas reservoir is exhausted or stellar feedback effects become important and remove the parental cocoon. Third, stellar feedback becomes so efficient that all  the remaining gas is cleared. This reveals the young cluster to optical telescopes, and its subsequent secular evolution is then largely dominated by gravitational processes rather than by the complex competition between gravity and many other physical agents.

We begin our discussion with a definition of what we actually mean when talking about star clusters in Section \ref{sec:def}.  Then we present evidence from analytic studies and numerical models that indicate that the proto-cluster gas is heavily influenced by the initial conditions and the dynamical properties of the parental cloud in Section \ref{sec:frag}.

As  gas contracts to form stars, the density increases by more than 25 orders of magnitude and the temperature rises by a factor of a million. The process comes to an end when nuclear burning processes set in and provide stability: a star is born. We discuss the different models and suggestions to describe this evolutionary phase in Section \ref{sec:sf}}. 

Star formation is controlled by the intricate interplay between the self-gravity of the ISM and various opposing agents, such as supersonic turbulence, magnetic fields, radiation and gas pressure. The evolution is modified by the thermodynamic response of the gas, which is determined by the balance between heating and cooling, which in turn depends on the chemical composition of the material and the detailed interaction of gas and dust with the interstellar radiation field. Altogether, stellar feedback provides enough energy and momentum to remove the parental gas from the cluster. It may also be responsible for clusters being in a global outflow state for the rest of their life. The various physical agents contributing to this process are discussed in Section \ref{sec:gd}. 

Once gas is removed, the subsequent dynamical evolution of a star cluster becomes relatively simple. It is solely governed by the mutual gravitational attraction of the stars in the cluster, modified only by tidal forces exerted from the larger-scale galactic environment, which are weak except near the galactic center or when clusters pass nearby overdensities (such as giant molecular clouds, GMCs, spiral arms, or other clusters), and by mass loss due to the internal evolution of the constituent stars. Large self-gravitating systems such as star clusters exhibit complex dynamical behavior which we discuss in Section  \ref{sec:nbody}.

The chemical composition of stars can provide important constraints on the origin of stellar clusters and help us to distinguish between different physical scenarios. We therefore introduce in Section \ref{sec:nucsy} key aspects of stellar nucleosynthesis and discuss their relation to cluster formation and evolution. Specifically, we speculate about the physical reasons for the observed O-Na anti-correlation observed in globular clusters. 

These different perspectives emphasise the interdependence of the different processes: How long gas remains in a state of turbulence before accreting onto a star (\S\ref{sec:frag}), and how accretion discs are connected to the upper hierarchies of the gas structure (\S\ref{sec:sf}) is crucial to understand how, and what kind of massive stars can pollute the gas out of which the low-mass stars form in massive star clusters and why this is not happening in lower-mass clusters (\S\ref{sec:nucsy}). Stellar feedback determines the abundance of gas in the cluster at all times after formation (\S\ref{sec:gd}) with implications for star formation (\S\ref{sec:sf},\S\ref{sec:nucsy}) and the dynamics of the stars (\S\ref{sec:nbody}).  

In summary, star clusters originate from a large reservoir of diffuse gas and dust that permeates the Galaxy, the interstellar medium. The process is governed by the complex interplay of often competing physical agents such as gravity, turbulence, magnetic fields, and radiation across the entire electromagnetic wavelength range. The system is organised in a hierarchy of scales, that link the global dynamical evolution of the galactic gas, to dense, star-forming clouds, and eventually to the newly born star clusters in their interior. Stellar feedback creates highly non-linear feedback loops that strongly influence the dynamical evolution across the entire cascade of scales. 

We provide an overview of the most important physical agents involved in the formation and early evolution of star clusters. We argue that stellar birth is a highly dynamical event, and that it couples a wide range of scales in space, time, and energy across the overall hierarchical structure of the galaxy. Star clusters are key constituents of the complex galactic ecosystem, where large parts evolve far from equilibrium and which exhibits highly non-linear dynamical behavior. Progress in this field rests on a comprehensive understanding of the underlying physics and chemistry. Due to the stochastic nature of problem, any theory of star formation is necessarily based on a statistical approach combined with an inventory of the different Galactic environments and knowledge of their possible variations across all scales. This is the direction in with future research efforts in this field will go.


\begin{acknowledgements}
This review emerged from a workshop at the International Space Science Institute in Bern, Switzerland in May 2019. The authors thank the staff of ISSI for their generous
hospitality and creating the inspiring atmosphere that initiated this project.
We thank the referee for a constructive report which has improved the manuscript.

\end{acknowledgements}

%
\section*{Conflict of interest}
J.M.D.K. gratefully acknowledges funding from the German Research Foundation (DFG) in the form of an Emmy Noether Research Group (grant number KR4801/1-1) and a DFG Sachbeihilfe Grant (grant number KR4801/2-1), from the European Research Council (ERC) under the European Union's Horizon 2020 research and innovation programme via the ERC Starting Grant MUSTANG (grant agreement number 714907), and from Sonderforschungsbereich SFB 881 ``The Milky Way System'' (subproject B2) of the DFG. S.S.R.O. acknowledges funding from NSF Career grant AST-1650486. P.G. acknowledges funding from the European Research Council under ERC-CoG grant CRAGSMAN-646955. 

R.S.K.\ acknowledges support from the Deutsche Forschungsgemeinschaft via the SFB 881 “The Milky Way System” (subprojects B1, B2, and B8) as well as funding from the Heidelberg Cluster of Excellence STRUCTURES in the framework of Germany’s Excellence Strategy (grant EXC-2181/1 - 390900948). 

JBP acknowledges UNAM-DGAPA-PAPIIT support through grant number IN-111-219. 

MG acknowledges support from the ERC (CLUSTERS, StG-335936).

E.V.-S. acknowledges financial support from CONACYT grant 255295.

\bibliographystyle{mnras}
\bibliography{references,bib2}   

\def\CCquestion#1{{\color{magenta} \sf #1}}
\def\CC#1{{\color{blue} \sf #1}}

\def\CCquestion#1{{\color{magenta} \sf #1}}
\def\CC#1{{\color{blue} \sf #1}}

\end{document}

%% file: def.tex
 \newcommand{\aap}{A\&A}
 \newcommand{\aapr}{A\&A Review}
 \newcommand{\aaps}{A\&A Supplement series}
 \newcommand{\mnras}{MNRAS}
 \newcommand{\apjs}{ApJS}
 
 \newcommand{\apj}{ApJ}
 \newcommand{\araa}{ARA\&A}
 \newcommand{\memsai}{MmSAI}
 \newcommand{\aj}{AJ}
 \newcommand{\prl}{Phys. Rev. Lett.}
 \newcommand{\pre}{Phys. Rev. E}
 
 \newcommand{\apjl}{ApJ}
 \newcommand{\nat}{Nature}
 
 \newcommand{\pasp}{PASP}
 
 \newcommand{\pasj}{PASJ}
 \newcommand{\na}{New Astronomy}
 \newcommand{\bain}{Bulletin of the Astronomical Institutes of the Netherlands}
 \newcommand{\doi}[1]{\textsc{doi}: \href{http://dx.doi.org/#1}{\nolinkurl{#1}}}
 
\newcommand{\nbr}[1]{\left( #1 \right)} 
\newcommand{\sbr}[1]{\left[ #1 \right]} 

\newcommand{\feh}{\sbr{\mathrm{Fe/H}}}
\newcommand{\kms}{km~s$^{-1}$}
\newcommand{\psc}{cm$^{-2}$}

\newcommand{\epsff}{\epsilon_{\rm ff}}
\newcommand{\Mc} {M_{\rm c}}
\newcommand{\MJ} {M_{\rm J}}

\def\pcc{\ifmmode {{\rm cm}^{-3}}\else cm$^{-3}$\fi}